%% file: main.tex
\ifdefined\pdftexversion\pdfoutput=1\fi
\documentclass[sigconf,nonacm]{acmart}
\renewcommand\footnotetextcopyrightpermission[1]{}
\usepackage{booktabs}
\usepackage{array}
\usepackage{tikz}
\usetikzlibrary{positioning,arrows.meta}
\input{numbers}

\begin{document}

\title{Escalate or Close: How Fast-Tier Models Handle Held-Out Web Attack Families in Two-Tier Triage}

\author{Maciej Kisielewicz}
\email{maciej.kisielewicz@pwr.edu.pl}
\affiliation{%
  \department{Department of Applied Informatics}
  \institution{Wroc{\l}aw University of Science and Technology}
  \city{Wroc{\l}aw}
  \country{Poland}}

\author{Micha{\l} K{\k{e}}dziora}
\affiliation{%
  \department{Department of Applied Informatics}
  \institution{Wroc{\l}aw University of Science and Technology}
  \city{Wroc{\l}aw}
  \country{Poland}}

\author{Ngoc Thanh Nguyen}
\affiliation{%
  \department{Department of Applied Informatics}
  \institution{Wroc{\l}aw University of Science and Technology}
  \city{Wroc{\l}aw}
  \country{Poland}}

\renewcommand{\shortauthors}{Kisielewicz, K{\k{e}}dziora and Nguyen}

\input{sections/abstract}

\keywords{two-tier triage, web attack detection, held-out attack families, decision models,
calibration, conformal prediction}

\maketitle

\input{body}

\end{document}

%% file: numbers.tex
\newcommand{\NumAdvClefAttempted}{79}
\newcommand{\NumAdvClefBudget}{150}

\newcommand{\NumAdvClefCut}{4}

\newcommand{\NumAdvClefN}{72}
\newcommand{\NumAdvClefPastOne}{75\%}
\newcommand{\NumAdvClefPastTenth}{97\%}
\newcommand{\NumAdvClefQueries}{9{,}500}

\newcommand{\NumAdvClefStartOne}{68}
\newcommand{\NumAdvClefStartTenth}{60}
\newcommand{\NumAdvClefSuccess}{26\%}

\newcommand{\NumAdvClefflashMedianQ}{7}

\newcommand{\NumAdvJevAtHundred}{19\%}

\newcommand{\NumAdvJevAttempted}{100}
\newcommand{\NumAdvJevBudget}{300}

\newcommand{\NumAdvJevFinalPMax}{0.22}
\newcommand{\NumAdvJevFinalPMin}{0.16}
\newcommand{\NumAdvJevHighStart}{19\%}

\newcommand{\NumAdvJevLowStart}{56\%}

\newcommand{\NumAdvJevMidStart}{35\%}
\newcommand{\NumAdvJevN}{99}
\newcommand{\NumAdvJevPastOne}{71\%}

\newcommand{\NumAdvJevQueries}{22{,}300}
\newcommand{\NumAdvJevSameClef}{28\%}

\newcommand{\NumAdvJevStartOne}{98}
\newcommand{\NumAdvJevStartTenth}{89}
\newcommand{\NumAdvJevSuccess}{43\%}

\newcommand{\NumAdvTransferJevClef}{5\%}

\newcommand{\NumAnomalyEscMax}{94\%}
\newcommand{\NumAnomalyEscMin}{46\%}

\newcommand{\NumApiRoundTrip}{205}

\newcommand{\NumApiRpsOne}{3.4}
\newcommand{\NumApiRpsSixteen}{58.3}

\newcommand{\NumBatClefClean}{98\%}

\newcommand{\NumBatClefEnc}{94\%}

\newcommand{\NumBatClefInstr}{100\%}
\newcommand{\NumBatClefInstrBenign}{99\%}

\newcommand{\NumBatClefPadBenign}{97\%}

\newcommand{\NumBatClefflashInstrBenign}{24\%}

\newcommand{\NumBatCrsoldonePadBenign}{5\%}

\newcommand{\NumBatCrsoneClean}{95\%}

\newcommand{\NumBatCrsoneEnc}{91\%}

\newcommand{\NumBatJevAttackMin}{98.8\%}

\newcommand{\NumBatJevCleanOne}{99.4\%}

\newcommand{\NumBatJevInstrBenign}{100\%}

\newcommand{\NumBatJevShiftAbsMax}{0.026}

\newcommand{\NumBatModernbertBenignMin}{49.9\%}
\newcommand{\NumBatModernbertClean}{93\%}

\newcommand{\NumBatModernbertPad}{43\%}
\newcommand{\NumBatModernbertPadBenign}{50\%}

\newcommand{\NumBatQwenClean}{88\%}

\newcommand{\NumBatQwenNote}{23\%}

\newcommand{\NumBatQwenValue}{24\%}

\newcommand{\NumBatTfidfClean}{58\%}

\newcommand{\NumBatTfidfPad}{15\%}

\newcommand{\NumBatteryAttacks}{704}
\newcommand{\NumBatteryBenign}{300}

\newcommand{\NumBudModernbertTraversalOne}{35\%}

\newcommand{\NumCanaryAliasSwitch}{27 October}

\newcommand{\NumCanaryFirst}{6 October}
\newcommand{\NumCanaryLast}{3 November}
\newcommand{\NumCanaryMeanDiff}{0.010}
\newcommand{\NumCanaryN}{500}

\newcommand{\NumCascadeBenignMax}{5.5\%}
\newcommand{\NumCascadeBenignMin}{4.2\%}

\newcommand{\NumCascadeCrsoneFwdTwentyCalls}{20\%}
\newcommand{\NumCascadeCrsoneFwdTwentyMin}{89\%}
\newcommand{\NumCascadeCrsoneFwdTwentyTraversal}{97\%}
\newcommand{\NumCascadeCveAuthEsc}{29\%}
\newcommand{\NumCascadeCveEsc}{73\%}

\newcommand{\NumCascadeModernbertFwdTwentyTraversal}{79\%}

\newcommand{\NumCascadeTfidfFwdTwentyTraversal}{19\%}
\newcommand{\NumClefCalls}{55{,}842}

\newcommand{\NumClefClosedLogshell}{3\%}

\newcommand{\NumClefCostPerMillion}{168.35}
\newcommand{\NumClefCostPerYear}{614{,}000}

\newcommand{\NumClefCveClosed}{7\%}

\newcommand{\NumClefCveClosedAuth}{47\%}

\newcommand{\NumClefCveEsc}{97\%}

\newcommand{\NumClefDriftN}{200}
\newcommand{\NumClefEcmlBenignAboveHalf}{86.9\%}

\newcommand{\NumClefEcmlEscLdap}{15\%}

\newcommand{\NumClefEscFloor}{97.6\%}
\newcommand{\NumClefEscLogshell}{99\%}

\newcommand{\NumClefFlood}{122}
\newcommand{\NumClefHardEsc}{8.2\%}

\newcommand{\NumClefLatMedian}{375}

\newcommand{\NumClefLatNinetyNine}{1{,}103}

\newcommand{\NumClefPOneAbs}{0.026}

\newcommand{\NumClefRoundTrip}{17}

\newcommand{\NumClefTokensMean}{701}

\newcommand{\NumClefflashCostPerMillion}{26.66}

\newcommand{\NumClefflashEcmlEscLdap}{14\%}

\newcommand{\NumClefflashEscMax}{97\%}
\newcommand{\NumClefflashEscMin}{68\%}

\newcommand{\NumClefflashKnownClosed}{5.9\%}
\newcommand{\NumClefflashKnownClosedMax}{6.5\%}
\newcommand{\NumClefflashKnownClosedMin}{5.4\%}
\newcommand{\NumClefflashKnownClosedPMax}{0.27}
\newcommand{\NumClefflashKnownClosedPMin}{0.02}
\newcommand{\NumClefflashLatMedian}{372}

\newcommand{\NumClefflashLatNinetyNine}{936}

\newcommand{\NumClefflashNovelAurocMax}{0.996}
\newcommand{\NumClefflashNovelAurocMin}{0.895}

\newcommand{\NumClusterModernbertPTwoHi}{+23.2}
\newcommand{\NumClusterModernbertPTwoLo}{+18.9}

\newcommand{\NumConfLoadClef}{1.5\%}
\newcommand{\NumConfLoadClefflash}{32.6\%}

\newcommand{\NumConfLoadJev}{0.9\%}

\newcommand{\NumCostBreakFive}{20}
\newcommand{\NumCostBreakFiveUsd}{\$0.0005}
\newcommand{\NumCostBreakOne}{100}
\newcommand{\NumCostBreakOneUsd}{\$0.003}
\newcommand{\NumCostBreakTenth}{1{,}000}
\newcommand{\NumCostBreakTenthUsd}{\$0.03}
\newcommand{\NumCrsCveAuthZero}{80\%}

\newcommand{\NumCrsoldoneDefBenign}{2.4\%}

\newcommand{\NumCrsoldoneDefHeldMax}{100\%}
\newcommand{\NumCrsoldoneDefHeldMin}{81\%}

\newcommand{\NumCrsoldoneDefLogshellParam}{100\%}
\newcommand{\NumCrsoldoneDefLogshellUa}{2\%}

\newcommand{\NumCrsoneDefBenign}{7.1\%}

\newcommand{\NumCrsoneDefCve}{70\%}
\newcommand{\NumCrsoneDefHdrBenign}{8.1\%}
\newcommand{\NumCrsoneDefHeldMax}{100\%}
\newcommand{\NumCrsoneDefHeldMin}{87\%}

\newcommand{\NumCrsoneEscMax}{100\%}
\newcommand{\NumCrsoneEscMin}{81\%}

\newcommand{\NumCrstwoDefBenign}{32.5\%}

\newcommand{\NumCsicBenignCal}{3{,}549}
\newcommand{\NumCsicBenignInScope}{18{,}620}
\newcommand{\NumCsicBenignTest}{3{,}615}

\newcommand{\NumCsicClefEscFloor}{99.8\%}

\newcommand{\NumCsicClefflashEscFloor}{93\%}

\newcommand{\NumCsicModernbertConfMissTraversal}{25\%}

\newcommand{\NumCsicQwenBenignAboveHalf}{32.4\%}

\newcommand{\NumCsicQwenEscMax}{54\%}
\newcommand{\NumCsicQwenEscMin}{29\%}

\newcommand{\NumCsicTfidfConfMissTraversal}{59\%}

\newcommand{\NumCveExclHeader}{126}

\newcommand{\NumCveHidden}{201}

\newcommand{\NumCveJevClosedPMax}{0.39}
\newcommand{\NumCveJevClosedPMin}{0.12}
\newcommand{\NumCveJevNotEscN}{9}
\newcommand{\NumCveLoaded}{624}
\newcommand{\NumCveMcpJevP}{0.17}

\newcommand{\NumCveNaiveJevClosed}{22\%}
\newcommand{\NumCveNaiveJevEsc}{88\%}
\newcommand{\NumCveNovelClassN}{114}
\newcommand{\NumCveRequests}{813}

\newcommand{\NumCveVisible}{423}

\newcommand{\NumDriftBenignOld}{5.2\%}
\newcommand{\NumDriftCanaryFlips}{19}
\newcommand{\NumDriftCanaryMeanDiff}{0.067}

\newcommand{\NumDriftEscNew}{98.4\%}
\newcommand{\NumDriftEscOld}{99.3\%}

\newcommand{\NumEcmlBenignQuote}{12\%}
\newcommand{\NumEcmlFamilyMax}{2{,}279}
\newcommand{\NumEcmlFamilyMin}{1{,}825}
\newcommand{\NumEcmlHeldFamilies}{six}
\newcommand{\NumEcmlHeldOutMax}{2{,}172}
\newcommand{\NumEcmlHeldOutMin}{1{,}745}

\newcommand{\NumEcmlMultiType}{2{,}180}
\newcommand{\NumEcmlSamples}{50{,}116}
\newcommand{\NumEcmlSpaceRunBenign}{0.1\%}
\newcommand{\NumEcmlSpaceRunFamilies}{four}
\newcommand{\NumEcmlSpaceRunMax}{99\%}
\newcommand{\NumEcmlSpaceRunMin}{70\%}
\newcommand{\NumEcmlSsiKept}{16}
\newcommand{\NumEcmlTfidfNovelAurocMax}{1.00}
\newcommand{\NumEcmlTfidfNovelAurocMin}{0.92}
\newcommand{\NumEcmlValid}{35{,}006}

\newcommand{\NumEncMbCmdexeDec}{97.8\%}

\newcommand{\NumEncMbTraversalDecCm}{61.6\%}

\newcommand{\NumEncMbTraversalOrigCm}{59.1\%}

\newcommand{\NumEncMbXssDec}{99.8\%}

\newcommand{\NumEncNgramCmdexeDec}{14\%}
\newcommand{\NumEncNgramCmdexeOrig}{86\%}
\newcommand{\NumEncNgramDecKnownMax}{97\%}
\newcommand{\NumEncNgramDecKnownMin}{87\%}
\newcommand{\NumEncNgramDecOtherMax}{61\%}
\newcommand{\NumEncNgramDecOtherMin}{8\%}

\newcommand{\NumEncNgramOrigOtherMax}{94\%}
\newcommand{\NumEncNgramOrigOtherMin}{36\%}

\newcommand{\NumEncNgramXssDec}{15\%}
\newcommand{\NumEncNgramXssOrig}{88\%}

\newcommand{\NumEndpointInTraining}{68\%}

\newcommand{\NumEnsConfMissTraversal}{38\%}
\newcommand{\NumEnsEscLogshell}{98\%}
\newcommand{\NumEnsEscTraversal}{75\%}

\newcommand{\NumEnsUnionEscTraversal}{77\%}
\newcommand{\NumEscParenBenign}{0.2\%}
\newcommand{\NumEscParenMax}{100\%}
\newcommand{\NumEscParenMin}{67\%}
\newcommand{\NumEscParenTraversal}{0\%}

\newcommand{\NumFloodClefAllMin}{8\%}

\newcommand{\NumFloodClefSelMin}{49\%}
\newcommand{\NumFloodModernbertAllMax}{23\%}
\newcommand{\NumFloodModernbertAllMin}{15\%}
\newcommand{\NumFloodModernbertSelMax}{33\%}
\newcommand{\NumFloodModernbertSelMin}{14\%}
\newcommand{\NumFloodPadClef}{10}
\newcommand{\NumFloodPadJev}{23}

\newcommand{\NumFloodSelectedN}{51}

\newcommand{\NumFormatSplicedAuroc}{0.496}
\newcommand{\NumFormatVendorAuroc}{1.000}
\newcommand{\NumFormsJevEscMin}{98.7\%}

\newcommand{\NumFormsPlainModernbertMiss}{76\%}

\newcommand{\NumHdrClefBenign}{4.3\%}

\newcommand{\NumHdrClefMin}{99.0\%}

\newcommand{\NumHdrCrsoneLogshellUa}{67\%}
\newcommand{\NumHdrCrsoneLogshellXapi}{66\%}

\newcommand{\NumHdrJevBenign}{4.4\%}

\newcommand{\NumHeaderBenign}{2{,}000}

\newcommand{\NumHostedDropMax}{5}
\newcommand{\NumHostedFamPTwoAbsMax}{1.2}
\newcommand{\NumHostsMinRequests}{50}
\newcommand{\NumHostsN}{21}
\newcommand{\NumJevCalls}{72{,}095}

\newcommand{\NumJevClosedLogshell}{2\%}
\newcommand{\NumJevClosedMax}{2.5\%}

\newcommand{\NumJevCostPerMillion}{25.59}
\newcommand{\NumJevCostPerYear}{93{,}000}
\newcommand{\NumJevCveAfterN}{4}

\newcommand{\NumJevCveAuthN}{35}
\newcommand{\NumJevCveClosed}{5\%}

\newcommand{\NumJevCveClosedAuth}{23\%}

\newcommand{\NumJevCveClosedHi}{7\%}
\newcommand{\NumJevCveClosedLo}{3\%}

\newcommand{\NumJevCveContentEsc}{97\%}
\newcommand{\NumJevCveEsc}{98\%}

\newcommand{\NumJevCveEscAuth}{86\%}

\newcommand{\NumJevCveEscHi}{99\%}
\newcommand{\NumJevCveEscLo}{96\%}

\newcommand{\NumJevCveTrendEsc}{$-2.5$}
\newcommand{\NumJevCveTrendEscP}{0.08}
\newcommand{\NumJevDeployAttacks}{10{,}000}
\newcommand{\NumJevDeployEscalations}{60{,}507}
\newcommand{\NumJevDeployFalseAlerts}{30{,}969}
\newcommand{\NumJevDeployMissed}{248}
\newcommand{\NumJevEcmlBenignAboveHalf}{91.8\%}
\newcommand{\NumJevEcmlBenignMedian}{0.80}
\newcommand{\NumJevEcmlEscLdap}{10\%}

\newcommand{\NumJevEscLogshell}{100\%}

\newcommand{\NumJevEscTraversal}{99\%}

\newcommand{\NumJevFlood}{99}

\newcommand{\NumJevHostEscMax}{88\%}

\newcommand{\NumJevHostZeroN}{11}

\newcommand{\NumJevLatMedian}{300}

\newcommand{\NumJevLatNinetyNine}{473}

\newcommand{\NumJevNovelAurocMin}{0.993}

\newcommand{\NumJevOaBenignAboveHalf}{0.6\%}
\newcommand{\NumJevOaBenignMedian}{0.07}

\newcommand{\NumJevPOneAbs}{0.019}

\newcommand{\NumJevQuestionAurocMin}{0.99}

\newcommand{\NumJevRepeatFlipMax}{1.6\%}

\newcommand{\NumJevRepeatN}{1{,}813}

\newcommand{\NumJevTokensMean}{609}
\newcommand{\NumKnownClosedMaxRest}{5.3\%}
\newcommand{\NumKnownClosedPMinRest}{0.11}

\newcommand{\NumKvhLayaftTraversalHeld}{31.1\%}
\newcommand{\NumKvhLayaftTraversalKnown}{99.3\%}

\newcommand{\NumKvhModernbertTraversalHeld}{55.4\%}
\newcommand{\NumKvhModernbertTraversalKnown}{99.8\%}

\newcommand{\NumKvhTfidfTraversalHeld}{2.1\%}
\newcommand{\NumKvhTfidfTraversalKnown}{97.3\%}

\newcommand{\NumLatModernbertMedian}{21}

\newcommand{\NumLayaChoiceAurocMax}{0.76}
\newcommand{\NumLayaChoiceAurocMin}{0.56}

\newcommand{\NumLayaKnownClosed}{6.5\%}
\newcommand{\NumLayaKnownClosedMax}{6.8\%}
\newcommand{\NumLayaKnownClosedMin}{5.5\%}
\newcommand{\NumLayaKnownClosedPMax}{0.25}
\newcommand{\NumLayaKnownClosedPMin}{0.002}

\newcommand{\NumLayaSpreadAucMax}{0.58}
\newcommand{\NumLayaSpreadAucMin}{0.57}
\newcommand{\NumLayaYesnoAurocMax}{0.71}
\newcommand{\NumLayaYesnoAurocMin}{0.56}

\newcommand{\NumLayaftClosedLogshell}{17\%}

\newcommand{\NumLayaftConfMissTraversal}{76\%}

\newcommand{\NumLayaftNovelAurocTraversal}{0.739}

\newcommand{\NumLengthEscMax}{5}
\newcommand{\NumLongAttackMax}{2.9\%}
\newcommand{\NumLongAttackMin}{0.0\%}
\newcommand{\NumLongAttackXxe}{26\%}

\newcommand{\NumMbCkptEscTraversal}{81\%}
\newcommand{\NumMbRunConfMissLogshellMax}{51\%}

\newcommand{\NumMbRunConfMissLogshellMin}{12\%}
\newcommand{\NumMbRunConfMissTraversalMax}{60\%}

\newcommand{\NumMbRunConfMissTraversalMin}{37\%}

\newcommand{\NumMbRunEscTraversalMax}{81\%}

\newcommand{\NumMbRunEscTraversalMin}{50\%}

\newcommand{\NumMbRunKnownEscTraversalMax}{99.4\%}
\newcommand{\NumMbRunKnownEscTraversalMin}{98.4\%}
\newcommand{\NumMbRunsLogshell}{7}
\newcommand{\NumMbRunsTraversal}{10}

\newcommand{\NumModernbertClosedTraversal}{65\%}

\newcommand{\NumModernbertConfEscXss}{57\%}

\newcommand{\NumModernbertConfMissLogshell}{38\%}

\newcommand{\NumModernbertConfMissTraversal}{56\%}
\newcommand{\NumModernbertConfMissXss}{3\%}

\newcommand{\NumModernbertCveTrendEsc}{$-8.3$}
\newcommand{\NumModernbertCveTrendEscP}{0.01}

\newcommand{\NumModernbertDeployEscalations}{99{,}261}
\newcommand{\NumModernbertDeployFalseAlerts}{2{,}664}
\newcommand{\NumModernbertDeployMissed}{6{,}509}

\newcommand{\NumModernbertEcmlEscMin}{82.9\%}

\newcommand{\NumModernbertNearJev}{four}

\newcommand{\NumModernbertNovelAurocTraversal}{0.889}

\newcommand{\NumModernbertTravBothMax}{14\%}
\newcommand{\NumModernbertTravBothMin}{9\%}
\newcommand{\NumModernbertTravTauMax}{0.031}
\newcommand{\NumModernbertTravTauMin}{0.023}
\newcommand{\NumNestedClefEsc}{99\%}

\newcommand{\NumNestedJevEsc}{100\%}

\newcommand{\NumNovOtherMaxDiff}{4}
\newcommand{\NumNovTravEither}{87\%}
\newcommand{\NumNovTravProb}{81\%}
\newcommand{\NumOaBenignCal}{10{,}000}
\newcommand{\NumOaBenignTest}{10{,}000}
\newcommand{\NumOaBenignTrain}{221{,}634}
\newcommand{\NumOaDistinct}{773{,}363}
\newcommand{\NumOaHardBenign}{503}
\newcommand{\NumOaHardSearched}{64{,}068}
\newcommand{\NumOaHeldOutCmdexe}{1{,}162}
\newcommand{\NumOaHeldOutLogshell}{104}

\newcommand{\NumOaHeldOutShellshock}{23}
\newcommand{\NumOaHeldOutSqli}{427}
\newcommand{\NumOaHeldOutTraversal}{2{,}858}
\newcommand{\NumOaHeldOutXss}{2{,}839}
\newcommand{\NumOaHeldOutXxe}{35}
\newcommand{\NumOaHostsInScope}{6{,}259}
\newcommand{\NumOaInScope}{388{,}691}
\newcommand{\NumOaPayloadCap}{3{,}000}
\newcommand{\NumOaPayloadsCmdexe}{1{,}234}
\newcommand{\NumOaPayloadsLogshell}{110}
\newcommand{\NumOaPayloadsShellshock}{24}
\newcommand{\NumOaPayloadsSqli}{458}

\newcommand{\NumOaPayloadsXxe}{35}
\newcommand{\NumOaRequests}{1{,}040{,}242}
\newcommand{\NumOaSessions}{692}
\newcommand{\NumOaTraversalCapped}{3{,}000}

\newcommand{\NumOaTraversalNotUtfCapped}{1{,}718}

\newcommand{\NumOpenCveEscMax}{16\%}
\newcommand{\NumOpenCveEscMin}{7\%}
\newcommand{\NumOpenEcmlEscMax}{10\%}
\newcommand{\NumOpenEcmlEscMin}{5\%}
\newcommand{\NumOpenEscMax}{23\%}
\newcommand{\NumOpenNovelAurocMax}{0.80}
\newcommand{\NumOpenQuestionAurocMax}{0.77}

\newcommand{\NumPadLongerShare}{7.0\%}

\newcommand{\NumPadMinChars}{3{,}281}
\newcommand{\NumPoolCal}{73{,}491}
\newcommand{\NumPoolTest}{74{,}068}
\newcommand{\NumProAfterClefN}{5}

\newcommand{\NumProClefEscAfterClefK}{4}

\newcommand{\NumProJevEscLo}{66\%}

\newcommand{\NumProN}{9}

\newcommand{\NumQwenCveEsc}{82\%}

\newcommand{\NumQwenCveNovelClassEsc}{57\%}

\newcommand{\NumQwenCveTrendEsc}{$-6.9$}
\newcommand{\NumQwenCveTrendEscP}{0.04}

\newcommand{\NumQwenEcmlBenignAboveHalf}{99.5\%}

\newcommand{\NumQwenEcmlEscMax}{7\%}
\newcommand{\NumQwenEcmlEscMin}{6\%}

\newcommand{\NumRawClefEceHeld}{0.07}

\newcommand{\NumRawJevEceHeld}{0.08}

\newcommand{\NumRawRuleJevEcmlBenign}{91.8\%}

\newcommand{\NumRawRuleJevOaHeldMax}{100\%}
\newcommand{\NumRawRuleJevOaHeldMin}{96\%}

\newcommand{\NumRecalDecHeldMax}{1.3\%}
\newcommand{\NumRecalSupHeldMax}{33\%}
\newcommand{\NumRecalSupHeldMin}{20\%}

\newcommand{\NumReqJevAll}{4}

\newcommand{\NumReqJevEsc}{60\%}
\newcommand{\NumReqJevMajority}{18}
\newcommand{\NumReqJevMaxDiff}{0.21}

\newcommand{\NumReqJevN}{43}
\newcommand{\NumReqJevRepeats}{5}
\newcommand{\NumReqJevRobustShare}{18\%}

\newcommand{\NumRrNeutralMaxAbs}{5}

\newcommand{\NumRrRedrawMaxAbs}{1.7}

\newcommand{\NumShellshockEscMin}{91\%}

\newcommand{\NumSingleCostPerYear}{72{,}000}
\newcommand{\NumSingleMaxDiff}{0.10}

\newcommand{\NumSingleN}{200}

\newcommand{\NumSingleTokensFive}{608}
\newcommand{\NumSingleTokensOne}{470}

\newcommand{\NumSmallClefMax}{91\%}
\newcommand{\NumSmallClefMin}{68\%}

\newcommand{\NumSmallClefflashMax}{63\%}
\newcommand{\NumSmallClefflashMin}{8\%}

\newcommand{\NumSmallCrsoldoneLogshell}{11\%}

\newcommand{\NumSmallJevMax}{97\%}
\newcommand{\NumSmallJevMin}{82\%}

\newcommand{\NumSmallModernbertLogshell}{31\%}

\newcommand{\NumSmallModernbertTraversal}{6\%}

\newcommand{\NumSmallTfidfLogshell}{4\%}

\newcommand{\NumSmallTfidfTraversal}{0\%}

\newcommand{\NumSpreadAucBelowMax}{0.44}
\newcommand{\NumSpreadAucBelowMin}{0.23}
\newcommand{\NumSrbhDistinct}{542{,}991}

\newcommand{\NumSrbhOneClient}{503{,}747}

\newcommand{\NumSrbhOneClientNormalSigShare}{46.1\%}
\newcommand{\NumSrbhOtherNormalSigShare}{2.8\%}

\newcommand{\NumStructEsc}{93\%}

\newcommand{\NumTemporalModernbertNoOldTwin}{2\%}

\newcommand{\NumTemporalModernbertTwin}{89\%}

\newcommand{\NumTemporalOldExploits}{2{,}190}

\newcommand{\NumTemporalTfidfEsc}{94\%}
\newcommand{\NumTemporalTfidfEscAuth}{71\%}

\newcommand{\NumTemporalTfidfNoOldEsc}{34\%}

\newcommand{\NumTemporalTfidfSsrfNeutral}{63\%}
\newcommand{\NumTemporalTfidfSsrfOrig}{94\%}
\newcommand{\NumTemporalTfidfTwin}{52\%}

\newcommand{\NumTfidfClosedLogshell}{80\%}

\newcommand{\NumTfidfEcmlEscMin}{78.6\%}

\newcommand{\NumTfidfFlood}{30}
\newcommand{\NumTfidfHardEsc}{33.5\%}

\newcommand{\NumTfidfNovelAurocTraversal}{0.553}

\newcommand{\NumTiesCrsMax}{6.5\%}
\newcommand{\NumTiesLaya}{15\%}
\newcommand{\NumTiesMaxRest}{1.4\%}

\newcommand{\NumTruncModernbertAll}{1.2\%}
\newcommand{\NumTruncModernbertTraversal}{0.9\%}
\newcommand{\NumTruncVerdictAll}{2.0\%}

\newcommand{\NumTwinAucCrsone}{0.895}
\newcommand{\NumTwinAucJev}{1.000}

\newcommand{\NumTwinClef}{27\%}

\newcommand{\NumTwinJev}{37\%}

\newcommand{\NumTwinNetCrsone}{73}

\newcommand{\NumTwinNetJev}{63}

\newcommand{\NumTwinNetQwen}{82}

\newcommand{\NumTwinOthersMax}{2.6\%}

\newcommand{\NumTwinsN}{62}
\newcommand{\NumTzeroAurocSqli}{0.23}
\newcommand{\NumTzeroAurocXss}{0.19}
\newcommand{\NumTzeroGo}{0.80}
\newcommand{\NumUnionBestGainP}{0.50}
\newcommand{\NumUnionPairs}{three}

\newcommand{\NumVerdictLegitAurocMax}{0.64}
\newcommand{\NumVerdictLegitAurocMin}{0.57}

\newcommand{\NumWilsonFullLoLogshell}{96\%}
\newcommand{\NumWilsonFullLoShellshock}{85\%}
\newcommand{\NumWilsonFullLoSqli}{99\%}
\newcommand{\NumWilsonFullLoXxe}{90\%}

%% file: sections/abstract.tex
\begin{abstract}
In two-tier triage a fast model decides most web requests and escalates the rest to a slow tier, so
an attack the fast tier closes is never examined. We hold each of seven attack families out of
training, recalibration and thresholds, splice the payloads into parameters of real browsing
traffic, and compare zero-shot decision models, which return a probability for a typed question,
with supervised detectors, a local language model, the OWASP Core Rule Set and an anomaly baseline.
Detectors trained on the known families can give much of a held-out family a confident benign
verdict: over \NumMbRunsTraversal{} training runs, ModernBERT gives \NumMbRunConfMissTraversalMin{}
to \NumMbRunConfMissTraversalMax{} of held-out path traversal a probability below 0.05 and escalates
only \NumMbRunEscTraversalMin{} to \NumMbRunEscTraversalMax{} of it within a 5\% benign budget,
against nearly all of the same requests when traversal is known. Jev and Clef, two hosted decision models whose
budget decisions do not depend on the held-out family, escalate at least \NumClefEscFloor{} of every
family within the same budget, but their undisclosed training data may contain these public
families. Tests on recent exploits need controls for request shape: Jev escalates \NumTwinJev{} of
benign twins (exploit requests with the attack replaced by legitimate values), and detectors trained
on older exploits learn what scanner requests look like. A search guided by Jev's probability gets
\NumAdvJevSuccess{} of held-out attacks past Jev within \NumAdvJevBudget{} queries, but Jev's answers
vary: asked again, it escalates those requests in \NumReqJevEsc{} of its answers.
\end{abstract}

%% file: body.tex
\input{sections/introduction}
\input{sections/related}
\input{sections/background}
\input{sections/design}
\input{sections/results_trust}
\input{sections/results_escalation}
\input{sections/results_adversary}
\input{sections/results_operations}
\input{sections/discussion}
\input{sections/conclusion}

\bibliographystyle{ACM-Reference-Format}
\bibliography{refs}

\appendix
\input{sections/ethics}
\input{sections/openscience}
\input{sections/appendix_ecml}
\input{sections/appendix_pitfalls}
\input{sections/appendix_additional}

%% file: sections/introduction.tex
\section{Introduction}
\label{sec:intro}

Security operations centers commonly triage in tiers. Tier-1 analysts work through a stream of alerts, many
of them false positives, and escalate to more experienced tiers the ones they cannot rule
out~\cite{kersten2023structure,alahmadi2022falsepositives}. Recent proposals for automated triage
follow the same pattern: a scanner or a fast decision model screens the input, and a large language
model (LLM) reviews what it forwards~\cite{sun2026cascade,liu2026systemone,mu2026hydrojev}. A language model
can examine a suspicious request in depth, but it takes seconds and charges per
token~\cite{zhao2025aidr}, too slow and costly for every request of a busy web service.
In such a pipeline a fast tier sees all traffic and decides most of it alone, and a slow tier (a generative
model, an agentic analysis or a person) examines only what the fast tier escalates.

A zero-day attack exploits a vulnerability before its disclosure; Bilge and Dumitra{\c{s}} found
such attacks lasting 312 days on average, and disclosure raising the volume of attacks on a
vulnerability by up to five orders of magnitude~\cite{bilge2012zeroday}. Until it is refitted, a fast
tier may meet such an attack with nothing in its training data, recalibration or thresholds that
resembles it; for the fast tier, the attack then belongs to an unseen family. The pipeline catches
it only if the fast tier does not close it, and escalation has to stay within a budget on how much benign
traffic the slow tier can absorb.

Since September 2026 a new class of models has been offered as the fast tier. TypeSafe's Jev, Cloudflare's Clef and the open Laya,
von and Verdict read an input once and return typed answers, such as a yes or no with a probability,
within a few hundred
milliseconds~\cite{typesafe2026jev,chen2026clef,laya2026card,von2026card,verdict2026card}; we call
them decision models. TypeSafe calls them System One models and claims calibrated probabilities, a property a fast tier
needs on attacks unlike any it was fitted on. Studies of
these models measure accuracy and calibration on general and security
decisions~\cite{deusser2026jev,liu2026systemone,sharma2026typedbench}, place one in front of a
reasoning model~\cite{mu2026hydrojev,li2026jevjudge}, and steer them with injected
text~\cite{hu2026jevadvbench,li2026samejudge}. To our knowledge, no study evaluates them as the fast tier
of web triage on
attack families held out of everything a deployment fits, against the detectors and rule sets
operators run, within a fixed benign budget.

We hold out one attack family at a time from training, recalibration and every threshold,
and ask four questions. Are the fast tier's probabilities trustworthy on the held-out family (RQ1)?
Which routing rule gets held-out attacks to the slow tier within a benign budget (RQ2)? What can an
adversary achieve by editing an attack or by sending benign requests that the fast tier escalates
(RQ3)? What does each fast tier cost in latency, money and slow-tier load (RQ4)? We compare six decision models used zero-shot,
three of them hosted, with three supervised detectors trained on the known families (ModernBERT, a
fine-tuned Laya and an n-gram model), a local
language model read through its token probabilities, the OWASP Core Rule Set (CRS)~\cite{owasp2026crs} and
a benign-only anomaly baseline. Our main data are real browsing requests from the open-appsec WAF
comparison~\cite{openappsec2026waf} with public attack payloads spliced in, among them Log4Shell
(CVE-2021-44228) and Shellshock (CVE-2014-6271). Exploits for \NumCveVisible{} vulnerabilities from
2025 and 2026, converted from nuclei templates~\cite{projectdiscovery2026nuclei}, test recent attacks,
and those published after the models' release are unlikely to be in their pretraining data. Benign
twins (exploit requests with the attack replaced by a legitimate value) and re-rendered requests
control for request shape. ECML/PKDD
2007~\cite{ecml2007challenge} and CSIC 2010~\cite{csic2010dataset} serve as further datasets.

Our contributions are:
\begin{itemize}
\item \emph{Supervised detectors confidently miss some held-out families, by amounts that vary
between training runs} (Sections~\ref{sec:trust} and~\ref{sec:escalation}). Over \NumMbRunsTraversal{} runs,
ModernBERT gives \NumMbRunConfMissTraversalMin{} to \NumMbRunConfMissTraversalMax{} of held-out path
traversal a probability below 0.05, although it escalates nearly all of the same requests when
traversal is known.
The misses persist under other recalibration maps and on percent-decoded requests, while the n-gram
model's escalation of the other families rests largely on how the benchmark encodes payloads.
\item \emph{Used zero-shot, Jev and Clef escalate nearly all of every held-out family}
(Section~\ref{sec:escalation}). They escalate at least \NumClefEscFloor{} of every family within a
5\% benign budget, and at 0.1\% Jev escalates \NumSmallJevMin{} to \NumSmallJevMax{} and Clef
\NumSmallClefMin{} to \NumSmallClefMax. Their training data may contain these public families, but
Jev escalates all \NumProN{} exploits published after its release and Clef
\NumProClefEscAfterClefK{} of the \NumProAfterClefN{} published after its own.
\item \emph{Exploit tests need controls for request shape} (Section~\ref{sec:cve}). Jev escalates
\NumTwinJev{} of the benign twins of exploit requests, and detectors trained on older exploits
escalate half or more, so much of their gain from the older exploits reflects request shape.
\item \emph{Evasion tests need repeated calls when a fast tier's answers vary}
(Section~\ref{sec:adaptive}). A score-guided search gets \NumAdvJevSuccess{} of held-out attacks past
Jev, but, asked again, Jev escalates those requests in \NumReqJevEsc{} of its answers; Clef answers
identically when asked again, and the \NumAdvClefSuccess{} of attacks the search gets past it within
\NumAdvClefBudget{} queries stay past it.
\item \emph{Flooding, cost and cascades} (Sections~\ref{sec:flood} and~\ref{sec:operations}). Padding benign
requests makes Clef escalate \NumBatClefPadBenign{} of them, enough to flood the slow tier. At a 5\%
budget the slow tier costs more than Jev's fast tier once an examination costs more than
\NumCostBreakFive{} Jev calls. A cascade with the rule set first escalates at least
\NumCascadeCrsoneFwdTwentyMin{} of every held-out family but \NumCascadeCveEsc{} of the recent
exploits.
\end{itemize}
Appendix~\ref{sec:hygiene} documents five pitfalls in web attack data, among them a request template
that lets a classifier separate a benchmark's attacks from benign traffic without reading a payload.

%% file: sections/related.tex
\section{Related Work}
\label{sec:related}

\paragraph{Decision models.}
Preprints on decision models appeared within weeks of their release. Benchmarks measure their accuracy and calibration on general tasks:
Deu{\ss}er et al.\ find Jev's choice probabilities well calibrated but its yes-or-no probabilities in
need of a tuned threshold~\cite{deusser2026jev}, and TypedBench~\cite{sharma2026typedbench},
JEVal~\cite{duan2026jeval} and Sys1Cal-v1~\cite{porcedda2026sys1cal} measure calibration error,
sensitivity to wording and overstated probabilities. Swapping yes and no flips about half of Jev's
answers~\cite{rafe2026gates}, and the option labels of open decision models can override their
definitions~\cite{azizi2026labels}.
Cloudflare reports Clef's accuracy, latency and a forecasting Brier score, but no calibration-error
or reliability measurement~\cite{chen2026clef,cloudflare2026clefcard}. Closest to our setting,
HydroJEV places Jev as the first tier of SCADA alarm triage, ahead of a supervised classifier on
event subtypes absent from its labels~\cite{mu2026hydrojev}; JEV-as-a-Judge accepts Jev's confident verdicts and escalates the rest
to a reasoning judge~\cite{li2026jevjudge}; and Messoudi and Mansar choose risk-controlled thresholds
for decision scores~\cite{messoudi2026risk}. On security decisions, Liu finds that a low average
calibration error can hide attacks labeled safe with high confidence~\cite{liu2026systemone}, and
Cyber-Jev, a small model fine-tuned for HTTP-attack, phishing-URL and prompt-injection
checks, reports calibration that does not hold outside its training domain~\cite{subham2026cyberjev}.
Others attack these models: one appended opinion flips 12.1\% of the decisions of
\texttt{jev-1.13.0}~\cite{hu2026jevadvbench}, forged inputs bypass 19 to 29\% of the phishing emails
Jev flags but none of its moderation decisions~\cite{li2026samejudge}, and prompt injection shifts
its probabilities~\cite{wu2026hijacking,wang2026hiddenrisks}. These studies evaluate the models on fixed test sets or against crafted inputs. Our evaluation
differs in four ways. We hold a family out of recalibration and thresholds as well as training,
which exposes confident misses, the failure Liu describes~\cite{liu2026systemone}, in every
supervised detector and rarely in Jev or Clef (Section~\ref{sec:trust}). We compare every fast tier
within one benign budget (Section~\ref{sec:escalation}). Benign twins and a classifier that sees no
parameter value, the format control, measure how much of an exploit test's recall comes from request
shape (Section~\ref{sec:cve}). We judge a score-guided search's evasions on repeated calls
(Section~\ref{sec:adaptive}), which matters for a model whose answers vary.

\paragraph{Zero-day and unseen attacks.}
A learned detector meets a problem related to zero-day attacks~\cite{bilge2012zeroday}: attacks
unlike its training data. For web requests, Kruegel
and Vigna~\cite{kruegel2003webanomaly} and PAYL~\cite{wang2004payl} model normal parameters and
payloads, ZeroWall flags requests that an encoder-decoder trained on benign traffic cannot
reproduce~\cite{tang2020zerowall}, and RETSINA meta-learns unsupervised detectors across web
domains~\cite{li2023retsina}. CADE detects samples of families absent from
training~\cite{yang2021cade}, OWAD detects and adapts to a shift of normal
behavior~\cite{han2023owad}, and Sarhan et al.\ evaluate network intrusion detectors on attack
classes held out from training~\cite{sarhan2023zeroday}. Pretrained transformers have been
found to degrade less and to stay better calibrated out of
distribution~\cite{hendrycks2020pretrained,desai2020calibration}; our fine-tuned transformers rank
held-out families better than the n-gram model but are no better calibrated on them.

\paragraph{Evaluating learned detectors.}
Sommer and Paxson~\cite{sommer2010closedworld} explained why learned intrusion detectors struggle
outside the traffic they were built on, and Axelsson~\cite{axelsson2000baserate} showed that at
realistic base rates small false-alarm rates overwhelm analysts; our projections of slow-tier load
(Section~\ref{sec:baserate}) follow both.
Arp et al.~\cite{arp2022dosdonts} cataloged pitfalls such as spurious correlations, and
TESSERACT~\cite{pendlebury2019tesseract} showed how designs that ignore time inflate results; the
template shortcut of Section~\ref{sec:data} is a spurious correlation in a public web benchmark.
WAF-A-MoLE~\cite{demetrio2020wafamole} evaded learned web application firewalls by mutating
payloads.

\paragraph{Conformal evaluation and deferral.}
Transcend~\cite{jordaney2017transcend} and its successor~\cite{barbero2022transcending} used conformal
evaluation to decide when a malware classifier should no longer be trusted; a conformal guarantee
need not hold once exchangeability fails~\cite{barber2023beyond}, and Li shows that a conformal threshold
for intrusion detection loses its coverage under adversarial perturbation~\cite{li2026conformalids}.
Calibration and its decay under shift are well studied outside
security~\cite{guo2017calibration,ovadia2019shift}. Selective classification~\cite{geifman2017selective} and learning to
defer~\cite{madras2018defer,mozannar2020defer} train a model to abstain or hand over, and LLM
cascades escalate the queries a cheap model is unsure
about~\cite{chen2023frugalgpt,kotte2026ucci}, optimizing accuracy per unit of cost.

\paragraph{Prompt injection and language models in security operations.}
A model that reads attacker-controlled text can be steered by it: indirect prompt injection
compromises LLM-integrated applications~\cite{greshake2023indirect}, injection attacks have been
formalized and benchmarked~\cite{liu2024promptinjection}, and optimized injections deceive LLM
judges~\cite{shi2024judgedeceiver}. Surveys map language models in security and in security
operations centers~\cite{xu2024llmsecsurvey,habibzadeh2025socsurvey}. Recent systems triage alerts with
cooperating agents~\cite{wei2025cortex} or compress LLM reasoning to meet latency
budgets~\cite{zhao2025aidr}. Sun et al.\ place an LLM after a web scanner and report LLMs alone at
95.4 to 100\% recall and 49.6 to 85.0\% false positives~\cite{sun2026cascade}; they also evaluate how
deeply LLMs understand web payloads~\cite{sun2026payloadsem}. Confidence can be read from token
likelihoods~\cite{kadavath2022know} or verbal statements~\cite{tian2023justask}.

%% file: sections/background.tex
\section{Background}
\label{sec:background}

\subsection{Two-tier triage}

Every request goes first to the fast tier, which can close it as benign, raise an alert (report it
as an attack without the slow tier), or escalate it; only escalated requests reach the slow tier, an
arrangement that follows the dual-process account of judgment~\cite{kahneman2011thinking}, from
which TypeSafe takes the name of its System One models. For an attack
the fast tier does not recognize, the useful outcome is escalation, and because every benign request
escalated costs the slow tier time, we bound escalation by a benign budget, the share of benign
traffic the slow tier is prepared to see. The fast tier can sit inline, deciding before a request
reaches the application, or read a copy of the traffic, where a request it closes has already been
served and attacks are detected after the fact. Our escalation measurements apply to both and
bound what the pipeline detects, since the slow tier still has to recognize an escalated attack from
the same request the attacker wrote.

\subsection{Fast-tier models}

\begin{table*}[t]
\caption{The fast tiers we compare. Decision models answer a typed question with a probability in
one forward pass. Prices are per million input
tokens (output is not charged). Questions: how many of our five questions a model answers per request
(Section~\ref{sec:models}).}
\label{tab:models}
\footnotesize
\setlength{\tabcolsep}{4pt}
\begin{tabular}{llllllll}
\toprule
Model & Kind & Parameters & Access & Version & Released & Price & Questions \\
\midrule
Jev~\cite{typesafe2026jev,typesafe2026docs} & decision model & not disclosed & hosted (TypeSafe) &
  \texttt{jev-1.13.0} & 15 Sep 2026 & \$0.042 & 5, one call \\
Clef~\cite{chen2026clef,cloudflare2026clefmodel} & decision model & 27B (Qwen base) & hosted (Workers AI);
  open weights & model name only & 1 Oct 2026 & \$0.24 & 5, one call \\
Clef-flash~\cite{chen2026clef,cloudflare2026clefcard} & decision model & 9B (Qwen base) & hosted
  (Workers AI); open weights & model name only & 1 Oct 2026 & \$0.038 & 5, one call \\
Laya, von, Verdict~\cite{laya2026card,von2026card,verdict2026card} & decision models & 151M--421M
  & local & pinned revisions & 17--19 Sep 2026 & -- & 5 \\
Qwen2.5-7B~\cite{qwen2024qwen25} & language model & 7.6B (4-bit) & local (Ollama) & Q4\_K\_M &
  19 Sep 2024 & -- & 1 \\
ModernBERT~\cite{warner2024modernbert} & supervised & 149M & fine-tuned per fold & 3 seeds &
  base 19 Dec 2024 & -- & -- \\
Laya (ft)~\cite{laya2026card} & supervised & 421M & fine-tuned per fold & 1 seed & -- & -- &
  1 choice \\
n-gram model & supervised & 50,000 features & trained per fold & -- & -- & -- & -- \\
OWASP CRS~\cite{owasp2026crs} & rule-based WAF & rule set & ModSecurity v3 & 4.30.0 and 3.3.2 &
  2 Oct 2026; 30 Jun 2021 & -- & -- \\
Anomaly baseline & benign-only & per parameter & trained on benign traffic & -- & -- & -- & -- \\
\bottomrule
\end{tabular}
\end{table*}

Cloudflare's Clef and Clef-flash add a
decision head and low-rank adapters, trained with a Brier loss for calibration, to Qwen language
models of 27B and 9B parameters~\cite{chen2026clef}; their weights are public, but we query
Cloudflare's hosted service~\cite{cloudflare2026workersai}. Table~\ref{tab:models} lists the decision
models with the fast tiers an operator would otherwise deploy.

\subsection{Threat model}
\label{sec:threat}

We consider a web service behind a two-tier pipeline whose fast tier was fitted (trained,
recalibrated or given thresholds) on traffic that contains a set of known attack families, and three
adversaries.

The first holds an attack from a family on which the fast tier was never fitted and succeeds when
the fast tier closes it. We reproduce this adversary by holding the family out and measure the share
of held-out attacks that each routing rule closes, escalates or alerts on.

The second controls the rest of the attack request. It may add a parameter that asserts the
request is benign or instructs the classifier, append such text to the attacked value,
percent-encode the payload so that the server decodes it unchanged, pad the request so that the
payload falls beyond a model's input window, or carry the payload in a header. We test a fixed
battery of such edits and a search, guided by the target model's probability, over the same kinds of
edits (Section~\ref{sec:robustness}). Resubmitting a request until a nondeterministic model closes it
is out of scope, though each copy the model escalates reaches the slow tier.

The third wants to exhaust the slow tier. Under a budget rule every escalated request costs the slow
tier one examination, whatever it contains, so the adversary needs requests the fast tier escalates.
Blatant attacks would do, but they can be blocked or rate-limited at the edge, and so can requests
that carry a recognizable instruction to the classifier. Legitimate requests that merely look unusual
cannot be blocked this way, because real customers send them. We measure how often such attack-like
benign requests are escalated, which is also the false-escalation load they impose in normal
operation, and how often benign requests padded with opaque parameters are escalated.

Anyone can query a hosted model, or run an open one with our published questions, and tune an
attack against it, as our search does; a detector trained on an operator's own traffic stays
private.

%% file: sections/design.tex
\section{Methodology}
\label{sec:design}

\begin{figure}[t]
\centering
\input{figures/pipeline}
\Description{Schematic: a request enters the fast tier, which closes it, raises an alert, or
escalates it to the slow tier, a language model or an analyst. A dashed box below, joined to the fast
tier by a dotted line, lists what a deployment fits on benign traffic and the known families: for a
supervised detector, training, recalibration and thresholds; for a pretrained model used zero-shot,
only recalibration and thresholds. The held-out family never enters.}
\caption{The two-tier pipeline and what a deployment fits for each kind of fast tier. A closed
request reaches the application unexamined; only escalated requests reach the slow tier.}
\label{fig:pipeline}
\end{figure}
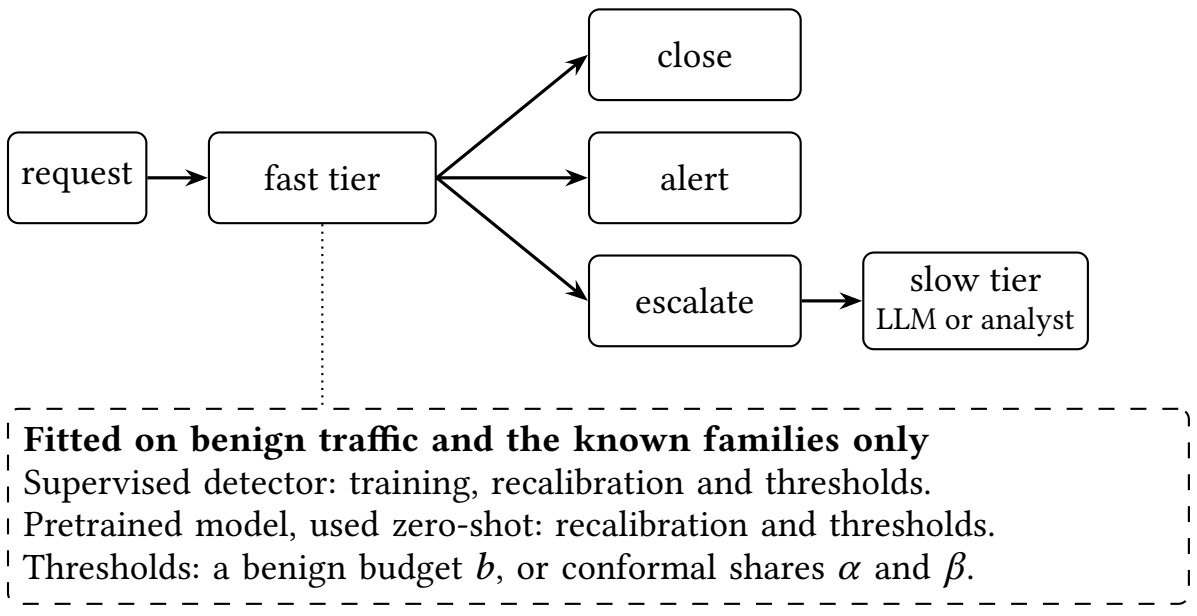

\subsection{Data}
\label{sec:data}

\paragraph{Benign traffic.}
For the 2026 edition of its WAF comparison, the open-appsec project published \NumOaRequests{} HTTP
requests that its authors recorded while browsing \NumOaSessions{} real websites in 14 categories and
using them as customers do: signing up, searching, filling carts and uploading
files~\cite{openappsec2026waf}. A browser records every request a page triggers, so the traffic
includes third-party trackers, content delivery networks and analytics alongside the sites themselves. We
represent each request in a compact form made of the request line, the Host and Content-Type
headers, and the body; the other headers come from the one recording browser or hold session state
such as cookies. After duplicates are removed, \NumOaDistinct{} distinct requests remain. We keep
those of at most 1,000 characters that carry at least one query, form or JSON-string parameter,
which leaves \NumOaInScope{} requests from \NumOaHostsInScope{} hosts. Every spliced attack
travels in a parameter, so without the parameter condition the mere absence of parameters would mark
a request as benign.

\paragraph{Attacks.}
The same project packages the mgm WAF Payload Collection~\cite{mgm2022payloads} into seven families
(Table~\ref{tab:data} in Appendix~\ref{sec:additional}). The project's harness sends every payload in the same request, a GET to \texttt{/?p=} or
a POST whose body is \texttt{p=}, so a classifier given only the method, Content-Type, Host, path and
parameter names, the \emph{format control}, separates these attacks from benign traffic with an AUROC of \NumFormatVendorAuroc,
without reading a single payload (Appendix~\ref{sec:hygiene}).
We splice each payload into one real request instead, replacing the value of one parameter, so that
attacks and benign traffic share hosts, paths and parameter names; the format control then reaches an
AUROC of \NumFormatSplicedAuroc.
Request and parameter are chosen by a hash of the payload, with no regard to what the parameter
means, so a traversal payload may land in a search field. Payloads enter query and form values
exactly as the harness transmits them, which matters for \NumOaTraversalNotUtfCapped{} of the
\NumOaTraversalCapped{} traversal payloads we keep, whose escapes, such as \texttt{\%c0\%ae}, decode
to bytes that are not valid UTF-8. A JSON string value receives the payload decoded once. In a
parameter value, a Shellshock or XML external entity payload tests only whether a fast tier
recognizes the string: CGI exports the query string whole, so no value reaches bash as a function
definition, and an entity acts only where an application parses the value as XML.
Section~\ref{sec:replay} also places Shellshock in headers, which a CGI server exports to bash. Spliced
attacks are not held to the benign length cap; leaving out the longer ones moves no model's held-out
escalation by more than \NumLengthEscMax{} points (Appendix~\ref{sec:additional}). We leave out the Padding
Evasion family that the 2026 edition adds, payloads behind kilobytes of junk, and test padding separately
(Section~\ref{sec:robustness}).

\paragraph{Attack-like benign traffic.}
Before running any model we fixed ten patterns that signature rules react to, such as markup,
quotes beside SQL keywords, \texttt{../}, template delimiters and shell separators before a command
name. We searched for them in the decoded parameter values of the \NumOaHardSearched{} benign
requests of the test split left out of the \NumOaBenignTest{} sampled for testing
(Table~\ref{tab:data}). The \NumOaHardBenign{} matches form an attack-like benign slice;
most come from ordinary page behavior, such as trackers that send \texttt{;cat=all} or template
macros like \texttt{\$\{TA\_DEVICE\_ID\}} that a page failed to expand.

\paragraph{ECML/PKDD 2007.}
For comparison with earlier work we also use the \NumEcmlSamples{} requests of the ECML/PKDD 2007
discovery challenge~\cite{ecml2007challenge}, \NumEcmlValid{} valid and the rest attacks of seven
curated types, \NumEcmlHeldFamilies{} of which we hold out in turn (Appendix~\ref{sec:ecml}).

\paragraph{CSIC 2010.}
As a second benign source we use the normal traffic of CSIC 2010~\cite{csic2010dataset}, synthetic
requests to one web shop, in the compact form and under the length and parameter rules of the main corpus, which
leaves \NumCsicBenignInScope{} distinct requests; the same payloads are spliced in the same way.

\paragraph{Exploits for recent vulnerabilities.}
The payload lists are public and old enough that they may have reached a pretrained model's
training data. As
a test less exposed to pretraining, we converted into requests the nuclei
templates~\cite{projectdiscovery2026nuclei} for vulnerabilities given a CVE identifier in 2025 or 2026,
keeping the \NumCveLoaded{}
templates tagged with an attack class such as SQL injection, file inclusion, server-side request
forgery, file upload or authentication bypass. An exploit's class is the first of its tags in a
fixed order; tags describe a vulnerability's impact as well as its payload, so the classes are
approximate. We evaluate each template's expressions as the scanner does, offline with random values
drawn from a fixed seed, and set the Host header to a host from the benign traffic. An exploit enters
the test when the compact form shows all of it and it carries attack content, which leaves
\NumCveVisible{} exploits with \NumCveRequests{} distinct requests. Of the other \NumCveHidden{}
exploits, \NumCveExclHeader{} carry part of the attack in a header the compact form drops, and the rest are
too long, cannot be reproduced offline, only request a page, or concern an earlier year. Of the
\NumCveVisible, \NumCveNovelClassN{} (file uploads, authentication bypasses and server-side request
forgeries) have no counterpart among the payload families. An exploit
counts as escalated when any of its requests is, and as closed when all of them are; the test
measures recall only. Thresholds come from the open-appsec folds, outcomes are averaged over the
seven folds, and NVD publication dates~\cite{nist2026nvdapi} place each exploit before or after a
model's release.

\subsection{Leave-one-family-out folds}
\label{sec:folds}

Each fold holds out one attack family, which appears only in the test split: training,
recalibration and every threshold see benign traffic and the remaining families. Before splitting we
set aside 5\% of the data, chosen by hash, as a development set, used only to fix the serialization,
to check the question wording on Jev, Laya, von and Verdict, the four decision models of the analysis
plan (Section~\ref{sec:metrics}), and von's controller setting (Appendix~\ref{sec:additional}), and
to screen a model trained before Log4Shell (Section~\ref{sec:limitations}); the wording, written before any run, did
not change. The remaining items are split 60/20/20 into training,
calibration and test by a hash of their text; for open-appsec an attack's split is decided by its
payload, so a payload never reaches two splits. Cross-site scripting, path traversal and command
execution, the three families with at least 1,000 held-out test attacks, form the main folds
(Table~\ref{tab:data}); SQL injection, Log4Shell, XML external entities and Shellshock are case
studies with Wilson intervals.

What a hold-out removes depends on the fast tier (Figure~\ref{fig:pipeline}). A supervised detector
never meets the family in training, recalibration or thresholds. A pretrained model is fitted only
through its recalibration map and thresholds; the map is monotone and the budget threshold is a
quantile of benign scores, so its budget-rule decision on a request is the same whichever family is
held out. Whether a family is new to a pretrained model depends on its pretraining, which the
providers do not document in full, so the hold-out tests what fitting to known families does to a fast
tier, and the exploit tests (Section~\ref{sec:robustness}) probe the pretrained models further.

\subsection{Model settings}
\label{sec:models}

\paragraph{Decision models.}
All six decision models answer the same frozen question set. The primary question is ``Does this
HTTP request attempt to attack, exploit, probe or abuse the web application?'' The set adds its
inverse (``Is this HTTP request ordinary, legitimate use of the web application?''), two paraphrases
(``Is this HTTP request malicious?'' and ``Would a security analyst flag this HTTP request as an
attack?'') and one choice among legitimate use, an attack, and a malformed but harmless request. A
model's attack probability is its probability of yes to the primary question (for von and Verdict,
see Appendix~\ref{sec:additional}). We left out
Decider and Bespoke Nimble, built on language models of about 2B and 9B
parameters~\cite{liu2026systemone}, a class Clef-flash and Qwen2.5-7B cover. We also left out
Cyber-Jev~\cite{subham2026cyberjev}, a classifier fine-tuned on web attacks that include our families,
which makes it a supervised detector.

\paragraph{Supervised detectors.}
The n-gram model is a logistic regression over TF-IDF weights of the 50,000 most frequent character
3- to 5-grams within whitespace-delimited tokens, trained on the full training split of each fold.
ModernBERT-base~\cite{warner2024modernbert} is fine-tuned per fold for one epoch on 8,000 benign
training requests chosen by hash and every known-family training attack (batches of 16, learning rate
$3\cdot10^{-5}$, at most 512 tokens, bf16), with seeds 17, 23 and 42, its \emph{main runs}. Training on the
GPU is not bit-for-bit reproducible, so even a repeated seed gives a new model; we trained ModernBERT
on the traversal and Log4Shell folds \NumMbRunsTraversal{} and \NumMbRunsLogshell{} times in all
(Section~\ref{sec:trust}). Analyses that score new texts (exploits, controls, manipulations, small
budgets) use one retraining of seed 17 whose checkpoints we kept, its \emph{kept run}; of the
\NumMbRunsTraversal{} runs on the traversal fold it escalates the most held-out traversal
(\NumMbCkptEscTraversal), so using it errs in ModernBERT's favor. Elsewhere ModernBERT's results on the
folds are the mean over its main runs unless we name a run. Laya (ft) is Laya fine-tuned per fold
with its own training loop~\cite{laya2026card}, for one epoch on ModernBERT's training selection with
one seed, and asked the primary question as a two-option choice with neutral option labels, as its card
recommends; it runs on open-appsec only. Nearly all payloads fit the input windows: the payload of \NumTruncModernbertAll{} of
held-out attacks (\NumTruncModernbertTraversal{} of traversal) ends beyond ModernBERT's 512 tokens.

\paragraph{Local language model.}
Qwen2.5-7B-Instruct~\cite{qwen2024qwen25} runs in Ollama 0.40.2 with Q4\_K\_M quantization, its chat template and default system prompt, at
temperature 0. It answers the primary question followed by ``Answer with one word: yes or no.'' Its
attack probability is P(yes)/(P(yes)+P(no)) over the first generated token, each word's probability
summing its case and spacing variants among the 20 most likely tokens.

\paragraph{Rule-based and benign-only baselines.}
The CRS~\cite{owasp2026crs}, the open rule set for ModSecurity, scores every request in ModSecurity
v3 in detection-only mode with all rules up to paranoia level 4. At the end of request processing, an
extra rule logs the inbound anomaly score each paranoia level contributes, so one pass gives the
score at levels 1 to 4. We use release 4.30.0 and release 3.3.2
of 30 June 2021, from before Log4Shell; ``the rule set'' means release 4.30.0 at paranoia level 1
unless we say otherwise. Every request is sent in its compact form with the same
fixed browser headers, and a request that nginx rejects before scoring gets the highest score. The anomaly baseline is fitted on benign training
requests only, after Kruegel and Vigna~\cite{kruegel2003webanomaly}: for every parameter name it
models the length of the decoded value (Chebyshev bound) and its character distribution (chi-square
against the idealized distribution), and a request scores the largest anomaly among its parameters.
Neither baseline is trained on attacks, so holding a family out changes neither's budget decisions.
The rule set's conformal rule takes its attack threshold from the known families' calibration
attacks and is reported with the other models'; the anomaly baseline is reported as escalation
within the benign budgets only.

\subsection{Recalibration and routing}
\label{sec:routing}

Every model's score is recalibrated per fold by monotone Platt
scaling~\cite{platt2000probabilistic,niculescu2005probabilities} on its logit, fitted on the
calibration split with the two classes weighted to equal mass; a negative fitted slope is replaced by
a constant, so an inverted model is reported as uninformative and never flipped. On the recalibrated
score we compare four routing rules.

\emph{Budget.} Escalate when the score exceeds the $(1-b)$-quantile of benign calibration scores, for
budgets $b$ of 1 and 5\% (and 0.1\% on the full benign pools, Section~\ref{sec:robustness}); a request
below it is closed. Ties, frequent in models that round or
saturate their outputs, are broken by a deterministic function of the request text, so a budget is
always spent exactly; at 5\% the tie-breaker settles at most \NumTiesMaxRest{} of test decisions,
except for Laya (\NumTiesLaya); the rule set, whose scores are integers (up to \NumTiesCrsMax); and
Verdict, whose recalibrated score is constant.

\emph{Conformal dual threshold.} Class-conditional conformal
sets~\cite{vovk2022alrw,angelopoulos2021gentle} admit the label attack when the score is at least
$t_{\mathrm{low}}$, the $k$-th smallest of the $n$ calibration attack scores, $k=\lfloor\alpha(n+1)\rfloor$,
and the label benign when it is at most $t_{\mathrm{high}}$, the mirror threshold on benign scores
with share $\beta$. A request is closed when only benign is admitted, alerted when only attack is,
and escalated when both or neither are; the empty set marks a score between the two classes'
thresholds. For attacks exchangeable with the calibration attacks, the rule closes at most a share
$\alpha$ in expectation; we use $\alpha=5\%$ and $\beta=1\%$. A held-out family is not exchangeable with the known
ones, so we measure the share of it the rule closes~\cite{barber2023beyond}.

\emph{Disagreement and combinations.} Escalate when either the score or the spread of the four
yes-or-no answers exceeds its threshold, or when either of two models' scores does; the two
thresholds share the budget so that together they escalate exactly the budget of benign calibration
traffic. A cascade instead lets a first stage (a supervised detector or the rule set) forward the
share of traffic it ranks highest to Jev, which decides on those requests.

The budget and conformal rules depend only on the order of the scores, so recalibration changes none
of their decisions, except where it reduces a model to a constant (Verdict); calibration matters for the confident-miss rate and when a model's own
probability is used as a threshold (Section~\ref{sec:trust}). When the known classes separate, so
that $t_{\mathrm{high}}<t_{\mathrm{low}}$, the conformal rule closes exactly the requests at or below
the benign $(1-\beta)$-quantile: the requests a budget of $\beta$ leaves unescalated. We verified this
equality on every such fold, which is every fold for Jev, ModernBERT and Laya (ft).

\subsection{Robustness and realism tests}
\label{sec:robustness}

\paragraph{Manipulation battery.}
We took \NumBatteryAttacks{} held-out attacks (200 from each main fold and all of Log4Shell, by hash)
and \NumBatteryBenign{} benign test requests, and applied five non-adaptive edits fixed before any
model scored them: a parameter asserting that the request is a routine health check and not an attack; a
parameter addressed to the classifier, asking it to answer that the request is legitimate; the same
assertion appended to the attacked value; every literal character of the payload percent-encoded,
which a server decodes to the same bytes; and about 3,000 characters of opaque parameters placed
before the payload. Each edited request is scored at its fold's clean thresholds, and ModernBERT's
kept run is compared with its own clean decisions.

\paragraph{Adaptive search.}
For \NumAdvJevAttempted{} of the battery's held-out attacks (30 each of traversal, cross-site
scripting and command execution, 10 of Log4Shell) a best-first random search in the manner of
WAF-A-MoLE~\cite{demetrio2020wafamole} keeps the five candidates the target scores lowest, mutates
one with one or two edits, and queries the target once per new candidate, until a candidate falls
below its fold's 5\% budget threshold or the queries run out. The edits keep the attack's effect
where added parameters go unread and the server decodes percent-encoding: added parameters (a benign assertion, an
instruction to the classifier or tracking-style padding); percent-encoding part of the payload; and,
per family, \texttt{./} segments in a path, changed letter case in HTML tag and attribute names,
empty quotes or carets in a shell word, and \texttt{\$\{lower:x\}} lookups in Log4Shell. The adversary reads only the
model's probability, as a public interface returns it. Jev, Clef-flash and Qwen2.5-7B get 300
queries per attack and Clef 150, on the first \NumAdvClefAttempted{} attacks of a rotation over the
families, where our spending cap for Clef ended the run; we leave out the \NumAdvClefCut{} searches
the cap cut short. Requests found against one model are then scored by the others, and those that
got past Jev or Clef are asked again \NumReqJevRepeats{} times.

\paragraph{Headers.}
Log4Shell reaches a vulnerable logger through any header an application logs, and Shellshock reaches
bash through headers a CGI server exports. We rebuilt every request in full from the recording, with the
browser's other headers but without cookies or credentials, and placed each held-out Log4Shell and
Shellshock payload, decoded once, three ways in its benign context: in the parameter as before, as
the User-Agent, and in an added X-Api-Version header. Thresholds come from \NumHeaderBenign{} benign
calibration requests in full form, and as many test requests give the realized benign rate.
Our supervised detectors read the compact form, which has no such headers, so in this test they
see neither header placement.

\paragraph{Small budgets.}
A 0.1\% budget would rest on the ten highest of the folds' \NumOaBenignCal{} calibration scores. We therefore scored every
benign request of the calibration and test splits, \NumPoolCal{} and \NumPoolTest, with the hosted
decision models, the n-gram model, ModernBERT's kept run, the rule set and the anomaly baseline, and
set a 0.1\% budget on the full calibration pool. Clef and Clef-flash, served without a version,
scored the rest of the pools twelve days after the folds; asked again, they answered
\NumClefDriftN{} fold requests identically.

\paragraph{Exploit tests and their controls.}
In a temporal test we trained ModernBERT and the n-gram model, as a deployment with older exploit
data would, on all seven families plus the attack-carrying requests of exploits for CVEs from 2015
to 2024 (\NumTemporalOldExploits{} exploits converted from the same nuclei snapshot with the same
rules, 80\% for training and 20\% for calibration by a hash of the CVE). An ablation used the seven
families alone; both are tested on the 2025--2026 exploits.
The exploits come from a scanner's templates, so they differ from browsing traffic in shape (admin
and API paths, random tokens, a callback host) as well as in attack content, and three controls
separate the two. \emph{Benign twins}: in \NumTwinsN{} exploit requests whose attack sits in a parameter value,
we replaced that value with a legitimate one (a search term, a file or user name, a same-site URL),
keeping everything else, the encoding included; uploads and most authentication bypasses, whose
attack is the file or the request itself, cannot be twinned. \emph{Re-rendering}: new random values
and callback subdomain, or neutral ones (word-like tokens and the callback under the request's own
host).
\emph{Shape only}: the format control of Section~\ref{sec:data}, which reads no parameter value,
trained and calibrated like the temporal test's detectors.

\paragraph{Prospective exploits.}
An exploit for a vulnerability that NVD published after a model's release is unlikely to be in that
model's pretraining data, though publication can follow first disclosure, and Clef may have changed since its release. From a second nuclei snapshot, of 23 October 2026~\cite{projectdiscovery2026nucleisecond}, we took the templates for
CVEs that NVD published after 15 September 2026, Jev's release, and kept those that pass the rules
above: \NumProN{} exploits, \NumProAfterClefN{} of them published after Clef's release on 1 October.

\subsection{Metrics and statistics}
\label{sec:metrics}

We read each test split as two slices, benign requests with known-family attacks and benign requests
with the held-out family, both weighted to equal class mass. Equal mass is a reference prior: at a
base rate of one attack in a thousand, $p=0.9$ at equal mass corresponds to a posterior of about
0.9\%, so routing is set by benign budgets. A confident miss is an attack given a probability below
0.05 at equal mass, which corresponds to a likelihood ratio below 1:19; the probability is the recalibrated one unless a
table gives the model's own.
Recalibrated scores are summarized by the square root of the debiased squared L2 calibration error of
Kumar et al.~\cite{kumar2019verified} over 15 equal-mass bins, which corrects for bin size; each
model's own probabilities by the L1 expected calibration error~\cite{naeini2015ece}, the quantity
calibration claims usually refer to. We report calibration error with the AUROC, because a constant
score has zero calibration error. Every escalation rate is reported with the realized escalation
of benign traffic.

Three comparisons are primary and confirmatory; every other analysis is exploratory. We fixed them in an internal, unregistered analysis plan
written before the first leave-one-family-out run; Clef, Clef-flash, Laya (ft) and ModernBERT's seeds
23 and 42 came after it. For every model but the two baselines, P1 is the calibration error on the held-out slice minus that on
the known slice, and P2 the same difference in confident-miss rate; for the six zero-shot decision
models, P3 compares held-out escalation at a 5\% budget under the disagreement rule against the budget rule.
P1 and P2 are pooled as the mean over main folds, with 95\% intervals from a stratified Bayesian
bootstrap~\cite{rubin1981bayesian} of 10,000 draws; P3 pools the paired decisions of all main folds
in an exact McNemar test. Holm's correction runs over the primary comparisons of the models in the
analysis plan and, separately, over those of the three models added after it; the intervals themselves are not adjusted. We also
report P2 per main family (Table~\ref{tab:primaryfamily}) and, because a family's attacks are often
variants of a few payloads, with a cluster bootstrap that gives one weight to all requests sharing a
payload skeleton (the payload decoded until stable, lower-cased, and every run of characters other
than letters and digits collapsed to one space).

%% file: figures/pipeline.tex
\begin{tikzpicture}[
    font=\footnotesize,
    box/.style={draw, rounded corners=2pt, align=center, inner sep=2.5pt, minimum height=6mm},
    outcome/.style={box, minimum width=14mm},
    arr/.style={-{Stealth[length=1.5mm]}, semithick}]
  \node[box] (req) {request};
  \node[box, right=4mm of req, minimum width=15mm] (fast) {fast tier};
  \node[outcome, right=10mm of fast] (alert) {alert};
  \node[outcome, above=2mm of alert] (close) {close};
  \node[outcome, below=2mm of alert] (esc) {escalate};
  \node[box, right=4mm of esc] (slow) {slow tier\\[-1pt]\scriptsize LLM or analyst};
  \draw[arr] (req) -- (fast);
  \draw[arr] (fast.east) -- (close.west);
  \draw[arr] (fast.east) -- (alert.west);
  \draw[arr] (fast.east) -- (esc.west);
  \draw[arr] (esc) -- (slow);
  \node[draw, dashed, rounded corners=2pt, align=left, inner sep=3pt, text width=76mm,
        anchor=north west, yshift=-4mm] (fit) at (req.west |- esc.south)
    {\textbf{Fitted on benign traffic and the known families only}\\
     Supervised detector: training, recalibration and thresholds.\\
     Pretrained model, used zero-shot: recalibration and thresholds.\\
     Thresholds: a benign budget $b$, or conformal shares $\alpha$ and $\beta$.};
  \draw[densely dotted] (fast.south) -- (fast.south |- fit.north);
\end{tikzpicture}

%% file: sections/results_trust.tex
\section{Calibration on Held-Out Attack Families (RQ1)}
\label{sec:trust}

\begin{table}[t]
\caption{Primary comparisons P1 and P2 (Section~\ref{sec:metrics}), main open-appsec folds:
held-out minus known calibration error and confident-miss rate (points) after recalibration, with
95\% intervals, and AUROC on known and held-out families. ModernBERT: main run, seed 17.
$^{\dagger}$: nonzero after Holm's correction; $^{\ddagger}$: added after the analysis plan, a Holm
family of its own.}
\label{tab:trust}
\footnotesize
\setlength{\tabcolsep}{3.5pt}
\input{tables/trust}
\end{table}

\begin{figure}[t]
\includegraphics[width=\columnwidth]{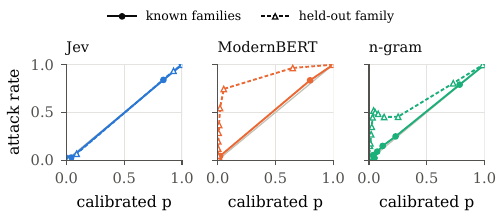}
\Description{Three small reliability plots, one per model, with recalibrated probability on the
horizontal axis and the observed share of attacks on the vertical axis, for known-family attacks
and for held-out attacks, next to the diagonal of perfect calibration.}
\caption{Observed attack rate against recalibrated probability, pooled over the main folds, for
known-family attacks (filled) and held-out attacks (hollow); ModernBERT: main run, seed 17. Points above the
diagonal at low probability are attacks the model calls benign with confidence.}
\label{fig:reliability}
\end{figure}

Routing by a benign budget uses only the order of the scores (Section~\ref{sec:routing}), but an
analyst or a fixed threshold reads the probability itself. On held-out families the low
probabilities of the supervised detectors cannot be trusted, and those of the hosted decision models
largely can, though for them the hold-out withholds a family only from recalibration and
thresholds.

\subsection{Supervised detectors miss with confidence}
\label{sec:trustfitted}
Recalibrated on the known families, as a deployment would be, the supervised detectors are calibrated
on those families and miscalibrated on held-out ones. Their calibration error and confident-miss
rate grow significantly (P1 and P2 in Table~\ref{tab:trust}); outside the tested comparisons, they
also grow on each main family alone, most on traversal (Table~\ref{tab:primaryfamily}), and by
similar amounts in ModernBERT's other two main runs (Table~\ref{tab:primary}). ModernBERT's main
runs separate held-out traversal from benign traffic with an AUROC of
\NumModernbertNovelAurocTraversal{} on average, yet give \NumModernbertConfMissTraversal{} of these
attacks a probability below 0.05, and \NumModernbertConfMissLogshell{} of held-out Log4Shell; on the
traversal fold their 5\% budget threshold lies below 0.05, so the budget rule still escalates some
of these confident misses (Appendix~\ref{sec:additional}). The share depends on the training run:
over \NumMbRunsTraversal{} runs that differ only in the seed or the GPU's nondeterminism, it ranges
from \NumMbRunConfMissTraversalMin{} to \NumMbRunConfMissTraversalMax{} for traversal, and over
\NumMbRunsLogshell{} runs from \NumMbRunConfMissLogshellMin{} to \NumMbRunConfMissLogshellMax{} for
Log4Shell (Table~\ref{tab:runs}), yet every run on the traversal fold escalates at least
\NumMbRunKnownEscTraversalMin{} of known-family attacks. Models that fit the training data equally
well can differ widely on data unlike it~\cite{damour2022underspecification}; an ensemble that
averages all runs' probabilities stays within these ranges on traversal
(Appendix~\ref{sec:additional}). A decision model fine-tuned on the known families fails the same
way: Laya (ft) ranks held-out traversal at an AUROC of \NumLayaftNovelAurocTraversal{} and gives
\NumLayaftConfMissTraversal{} of it a probability below 0.05, whereas zero-shot Laya gives none. The
n-gram model ranks held-out traversal barely above benign traffic (AUROC
\NumTfidfNovelAurocTraversal). Figure~\ref{fig:reliability} shows how the misses arise: on known
families the supervised detectors lie close to the diagonal, but on a held-out family many attacks
score in the range the recalibration map associates with benign traffic, so the map calls them
benign with high confidence even when they score above most benign requests.

\subsection{Other maps and decoded requests}
\label{sec:trustmap}
The misses do not depend on Platt scaling: with temperature scaling or isotonic regression, wherever
the map stays calibrated on the known families, the supervised detectors give \NumRecalSupHeldMin{} to
\NumRecalSupHeldMax{} of held-out attacks a probability below 0.05 and the hosted decision models at
most \NumRecalDecHeldMax{} (Table~\ref{tab:recal}). Nor do they come from how the benchmark encodes
payloads (Appendix~\ref{sec:hygiene}). Retrained with seed 17 on requests whose query and form values
are all percent-decoded, ModernBERT gives \NumEncMbTraversalDecCm{} of held-out traversal a
probability below 0.05, against \NumEncMbTraversalOrigCm{} for its seed-17 main run, and escalates at
least as much of held-out cross-site scripting and command execution as that run
(\NumEncMbXssDec{} and \NumEncMbCmdexeDec). The n-gram model does lean on the encoding: refitted the
same way, it still escalates \NumEncNgramDecKnownMin{} to \NumEncNgramDecKnownMax{} of known-family
attacks but only \NumEncNgramDecOtherMin{} to \NumEncNgramDecOtherMax{} of a held-out family other
than traversal, against \NumEncNgramOrigOtherMin{} to \NumEncNgramOrigOtherMax{} before.

\subsection{Hosted models stay nearly calibrated}
\label{sec:trustdecision}
Before recalibration, Jev's and Clef's own probabilities are as well calibrated on held-out families
as on known ones (expected calibration error \NumRawJevEceHeld{} and \NumRawClefEceHeld;
Table~\ref{tab:raw} in Appendix~\ref{sec:additional}). A fixed threshold of 0.5 does not carry over
between benign sources: Jev scores \NumRawRuleJevOaHeldMin{} to \NumRawRuleJevOaHeldMax{} of every
held-out family above it, but also \NumRawRuleJevEcmlBenign{} of the synthetic benign requests of
ECML/PKDD 2007 (Appendix~\ref{sec:ecml}).

Jev and Clef rank held-out families nearly as well as the known ones (AUROC at least
\NumJevNovelAurocMin{} for both on every fold); after recalibration their calibration errors grow by
\NumJevPOneAbs{} and \NumClefPOneAbs, significant but small, and their confident-miss rates by less
than half a point (Table~\ref{tab:trust}). Clef-flash ranks attacks less sharply (held-out AUROC
\NumClefflashNovelAurocMin{} to \NumClefflashNovelAurocMax), and neither its calibration error nor
that of Qwen2.5-7B changes significantly. Asked zero-shot, the small open models carry little
information about attacks: with the primary question Laya and von reach a held-out AUROC below
\NumOpenNovelAurocMax{} on every fold, and Verdict ranks attacks below benign traffic, so
recalibration reduces it to a constant (Table~\ref{tab:smallopen}).

%% file: tables/trust.tex
\begin{tabular}{lrrrr}
\toprule
 & \multicolumn{2}{c}{AUROC} & \multicolumn{2}{c}{Held-out $-$ known} \\
\cmidrule(lr){2-3}\cmidrule(lr){4-5}
Model & known & held & P1: calib.\ error & P2: conf.\ miss \\
\midrule
Jev & 0.998 & 0.997 & +0.019 [0.011, 0.023]$^{\dagger}$ & +0.4 [0.1, 0.8] \\
Clef$^{\ddagger}$ & 0.996 & 0.995 & +0.026 [0.018, 0.033]$^{\dagger}$ & +0.3 [$-$0.1, 0.7] \\
Clef-flash$^{\ddagger}$ & 0.931 & 0.933 & +0.012 [$-$0.003, 0.020] & $-$0.1 [$-$0.5, 0.3] \\
\midrule
Qwen2.5-7B & 0.971 & 0.971 & +0.008 [$-$0.001, 0.016] & +0.3 [$-$0.2, 0.8] \\
\midrule
ModernBERT & 0.998 & 0.959 & +0.217 [0.209, 0.228]$^{\dagger}$ & +20.8 [20.0, 21.6]$^{\dagger}$ \\
Laya (ft)$^{\ddagger}$ & 0.995 & 0.906 & +0.180 [0.170, 0.190]$^{\dagger}$ & +29.0 [28.0, 30.0]$^{\dagger}$ \\
n-gram model & 0.987 & 0.829 & +0.190 [0.177, 0.200]$^{\dagger}$ & +20.7 [19.8, 21.6]$^{\dagger}$ \\
\bottomrule
\end{tabular}

%% file: sections/results_escalation.tex
\section{Escalating Held-Out Attacks (RQ2)}
\label{sec:escalation}

\begin{table*}[t]
\caption{Share of held-out attacks (\%) escalated when each fast tier may escalate 5\% of benign
traffic, per held-out family; the realized share of benign test traffic escalated (Benign); and the
share of the \NumCveVisible{} recent exploits escalated (Section~\ref{sec:cve}). ModernBERT: mean over
its three main runs, exploits: its kept run (Section~\ref{sec:models}); --: not run on the exploits; CRS:
paranoia level 1. The last four families are case studies with fewer than 1,000
held-out attacks; for an observed 100\% the 95\% Wilson lower bound is \NumWilsonFullLoShellshock{}
(Shellshock), \NumWilsonFullLoXxe{} (XXE), \NumWilsonFullLoLogshell{} (Log4Shell) and
\NumWilsonFullLoSqli{} (SQL injection).}
\label{tab:escalation}
\footnotesize
\setlength{\tabcolsep}{5pt}
\input{tables/escalation}
\end{table*}

Under the plain budget rule at a 5\% budget, Jev and Clef escalate at least \NumClefEscFloor{} of every
held-out family while the supervised detectors lose much of traversal; at a 0.1\% budget Clef
escalates as little as \NumSmallClefMin{} of a family. A conformal rule's guarantee does not cover a
held-out family, and disagreement between questions or models never escalated significantly more
attacks. On recent exploits, much of what a recall test credits is request shape.

\subsection{Supervised detectors lose traversal}
\label{sec:budget}
For Jev and Clef, as for every zero-shot model, the hold-out changes no budget decision
(Section~\ref{sec:folds}). The other decision models escalate less within the 5\% budget
(Table~\ref{tab:escalation}): Clef-flash \NumClefflashEscMin{} to \NumClefflashEscMax{} of each family
and the small open models at most \NumOpenEscMax{} (Table~\ref{tab:smallopen}).

The supervised detectors lose held-out traversal, and the same requests show that holding traversal
out causes the loss. Of the traversal attacks in the other folds' test splits, where traversal is a
known family, ModernBERT escalates \NumKvhModernbertTraversalKnown, and
\NumKvhModernbertTraversalHeld{} once traversal is held out; Laya (ft) falls from
\NumKvhLayaftTraversalKnown{} to \NumKvhLayaftTraversalHeld{} and the n-gram model from
\NumKvhTfidfTraversalKnown{} to \NumKvhTfidfTraversalHeld. Over its \NumMbRunsTraversal{} training
runs ModernBERT escalates \NumMbRunEscTraversalMin{} to \NumMbRunEscTraversalMax{} of held-out
traversal, although it comes within two points of Jev on \NumModernbertNearJev{} of the seven
families. The fast tiers trained on no attack data cannot lose a family this way: at paranoia level 1
the rule set escalates \NumCrsoneEscMin{} to \NumCrsoneEscMax{} of each family and the anomaly
baseline \NumAnomalyEscMin{} to \NumAnomalyEscMax; higher paranoia levels mostly escalate less within
the budget (Table~\ref{tab:crslevels}).

With CSIC 2010 as the benign traffic, Jev and Clef escalate at least \NumCsicClefEscFloor{} of every
held-out family, and the supervised detectors again fall short on traversal, though less
(Table~\ref{tab:csic}).

\subsection{Smaller budgets}
\label{sec:small}
\begin{figure}[t]
\centering
\includegraphics[width=\columnwidth]{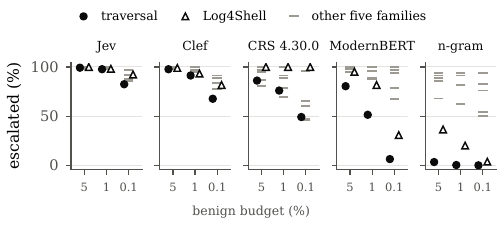}
\Description{Five small dot plots, one per fast tier (Jev, Clef, CRS 4.30.0, ModernBERT and the
n-gram model), of the share of held-out attacks escalated at benign budgets of 5\%, 1\% and 0.1\%: a
black dot for traversal, a hollow triangle for Log4Shell and gray ticks for the other five families.
Jev stays high for every family; Clef escalates about two thirds of traversal at the smallest budget.
The rule set keeps Log4Shell and cross-site scripting near the top and loses about half of the
others. ModernBERT loses most of traversal and Log4Shell at the smallest budget, and the n-gram model
escalates almost no traversal at any budget.}
\caption{Held-out attacks escalated at benign budgets of 5, 1 and 0.1\%, with thresholds set on all
\NumPoolCal{} benign calibration requests (Section~\ref{sec:robustness}). ModernBERT: its kept run
(Section~\ref{sec:models}).}
\label{fig:budgets}
\end{figure}
Analysts may not be able to examine 5\% of benign traffic. At a 0.1\% budget set on every benign
calibration request (Figure~\ref{fig:budgets}), Jev still escalates \NumSmallJevMin{} to
\NumSmallJevMax{} of each held-out family. Clef escalates only \NumSmallClefMin{} to
\NumSmallClefMax, least of traversal, and Clef-flash \NumSmallClefflashMin{} to
\NumSmallClefflashMax{} (Table~\ref{tab:small}). ModernBERT's kept run escalates
\NumSmallModernbertTraversal{} of traversal and \NumSmallModernbertLogshell{} of Log4Shell, and the
n-gram model \NumSmallTfidfTraversal{} and \NumSmallTfidfLogshell. Release 4.30.0 of the rule set
then escalates more held-out cross-site scripting and Log4Shell than Jev or Clef; release 3.3.2
escalates only \NumSmallCrsoldoneLogshell{} of Log4Shell, since most of these payloads score 5, the
weight of one critical rule, and the threshold now exceeds that score. With 1\% budgets set on the
calibration split, ModernBERT's main runs escalate \NumBudModernbertTraversalOne{} of traversal.

\subsection{No guarantee for a held-out family}
\label{sec:conformal}
\begin{table}[t]
\caption{Share of held-out attacks (\%) closed by the conformal rule fitted on the known families
($\alpha=5\%$, $\beta=1\%$); known: share of known-family test attacks closed; Load: benign test
traffic escalated or alerted (\%, main folds). $^{*}$: above 5\% with $p<0.001$ (exact binomial).}
\label{tab:conformal}
\footnotesize
\setlength{\tabcolsep}{2.5pt}
\input{tables/conformal}
\end{table}
On known families the conformal rule closes about the 5\% of attacks its guarantee allows or fewer,
with a small excess for Clef-flash (Table~\ref{tab:conformal}) and for Laya
(Appendix~\ref{sec:additional}). When the known classes separate, as for Jev, ModernBERT and Laya
(ft) on every fold, the rule closes exactly what a 1\% budget leaves unescalated
(Section~\ref{sec:routing}): Jev's rule closes at most \NumJevClosedMax{} of any held-out family, and
ModernBERT's \NumModernbertClosedTraversal{} of held-out traversal. The rule's benign load, escalated
or alerted, depends on how well the known classes separate, from \NumConfLoadJev{} of benign traffic
for Jev to \NumConfLoadClefflash{} for Clef-flash. Unlike the budget rule, the conformal rule
escalates a request whose score falls between the two classes' thresholds, an empty prediction set;
this is how ModernBERT escalates \NumModernbertConfEscXss{} of held-out cross-site scripting at a
benign load under 1\%.

\subsection{Disagreement and novelty signals}
\label{sec:signals}
The third primary comparison is negative. At an equal benign budget, escalating on confidence or on
the spread among the four yes-or-no answers caught fewer held-out attacks than confidence alone for
every zero-shot decision model (P3 in Table~\ref{tab:primary}). Across \NumUnionPairs{} pairs of
models and the seven held-out families, no union escalated significantly more attacks than its better
model alone (Appendix~\ref{sec:additional}). A novelty score helps ModernBERT on traversal: with the
smallest Mahalanobis distance of its embedding to its benign and known-family training
classes~\cite{lee2018mahalanobis} as a second signal sharing the budget, the kept run escalates
\NumNovTravEither{} of held-out traversal instead of \NumNovTravProb; no other family moves by more
than \NumNovOtherMaxDiff{} points (Table~\ref{tab:novelty} in Appendix~\ref{sec:additional}).

\subsection{Log4Shell and Shellshock, replayed}
\label{sec:replay}
With the other six families known, the supervised detectors' conformal rules close
\NumLayaftClosedLogshell{} (Laya (ft)) to \NumTfidfClosedLogshell{} (n-gram model) of held-out
Log4Shell, and those of Jev and Clef \NumJevClosedLogshell{} and \NumClefClosedLogshell{}
(Table~\ref{tab:conformal}). Within the budget Jev and Clef escalate \NumJevEscLogshell{} and
\NumClefEscLogshell{} of it, and \NumNestedJevEsc{} and \NumNestedClefEsc{} of the payloads that nest
the lookup, such as \texttt{\$\{\$\{::-j\}ndi:\ldots\}}. Shellshock is easier: the hosted models,
Qwen2.5-7B and the supervised detectors each escalate at least \NumShellshockEscMin{} of it, though in
a parameter it tests only recognition of the string (Section~\ref{sec:data}).

For the rule set the replay can also be faithful in time. Release 3.3.2, from June 2021, before
Log4Shell was disclosed in December, has no rule for JNDI lookups, but its level-1 rule for Unix shell
expressions (rule 932130) matches \texttt{\$\{\ldots\}} in parameters. As shipped, blocking at an
anomaly score of 5, it blocks \NumCrsoldoneDefLogshellParam{} of the held-out Log4Shell payloads
carried in a parameter, at \NumCrsoldoneDefBenign{} of benign requests. The nuclei template for
Log4Shell also places the payload in seventeen headers, User-Agent and X-Api-Version among
them~\cite{projectdiscovery2026nuclei}, and rule 932130 does not read headers: there the same rule
set blocks \NumCrsoldoneDefLogshellUa{} of the payloads, while release 4.30.0 blocks them in every
placement as shipped. In requests with headers, where Shellshock can reach bash
(Table~\ref{tab:headers} in Appendix~\ref{sec:additional}), Jev escalates every placement of both
families and Clef at least \NumHdrClefMin.

\subsection{Exploit tests need shape controls}
\label{sec:cve}
\begin{table}[t]
\caption{Controls for the exploit tests (Section~\ref{sec:robustness}): recent exploits escalated
(\%); \NumTwinsN{} exploit requests / their benign twins escalated (\%); change under a neutral
rendering (points); exploits published after 15 September 2026 escalated (of \NumProN, averaged over
folds and rounded). Rows below the line: trained with the 2015--2024 exploits as well; No values:
the format control. ModernBERT above the line: its kept run.}
\label{tab:controls}
\footnotesize
\setlength{\tabcolsep}{3pt}
\input{tables/controls}
\end{table}
Within the 5\% budget Jev escalates \NumJevCveEsc{} of the \NumCveVisible{} exploits and Clef
\NumClefCveEsc{} (Table~\ref{tab:escalation}; by attack class in Table~\ref{tab:cve}). Qwen2.5-7B,
released before every dated exploit~\cite{qwen2024blog}, escalates \NumQwenCveEsc, while the fast
tiers trained on no attack data do worse on the exploits than on the payload families. The benign
twins (Section~\ref{sec:robustness}) show how much of this is shape (Table~\ref{tab:controls}). Jev
escalates \NumTwinJev{} of the twins and Clef \NumTwinClef, against a 5\% budget, while the rule set,
Qwen2.5-7B and the supervised detectors escalate at most \NumTwinOthersMax. Jev scores every exploit
request above its twin (paired AUROC \NumTwinAucJev; rule set \NumTwinAucCrsone), but its threshold,
set on browsing traffic, lies below these scanner-shaped benign requests: net of its twins it
escalates \NumTwinNetJev{} points more of the exploit requests, fewer than the rule set
(\NumTwinNetCrsone) or Qwen2.5-7B (\NumTwinNetQwen). Rendering the exploits again changes little (at
most \NumRrRedrawMaxAbs{} points with new random values and \NumRrNeutralMaxAbs{} with neutral ones).

The controls also change what the temporal test shows. With the older exploits in training, the
n-gram model escalates \NumTemporalTfidfEsc{} of the recent exploits, against
\NumTemporalTfidfNoOldEsc{} in the ablation, and \NumTemporalTfidfEscAuth{} of the authentication
bypasses (Table~\ref{tab:temporalclasses}), but also \NumTemporalTfidfTwin{} of the benign twins
(none in the ablation). ModernBERT, trained the same way, escalates \NumTemporalModernbertTwin{} of the
twins (\NumTemporalModernbertNoOldTwin{} in the ablation), and the format control, which never sees a
parameter value, escalates \NumStructEsc{} of the recent exploits after the same training. These
detectors learned what scanner-generated exploit requests look like, which a test of recall alone
would credit as learning the attacks.

Of the exploits published after each model's release, Jev escalates all \NumProN{} (95\% exact lower
bound \NumProJevEscLo) and Clef \NumProClefEscAfterClefK{} of \NumProAfterClefN. On all recent
exploits their conformal rules close \NumJevCveClosed{} and \NumClefCveClosed{} on average over folds,
but not evenly: \NumJevCveClosedAuth{} and \NumClefCveClosedAuth{} of the \NumJevCveAuthN{}
authentication bypasses (exact binomial $p<0.001$ against 5\% for both). Many of the exploits Jev's
rule closes look like ordinary use of an application's interface (Table~\ref{tab:cveexamples} in
Appendix~\ref{sec:additional}): a token request or a password reset is hard to judge without the
application's state, and a harmless file sent to a vulnerable upload form carries no attack content.
One is a miss on visible attack content, an unauthenticated call to an agent server's
\texttt{terminal\_execute} tool with the command \texttt{id} (CVE-2026-59726), which Jev scores
\NumCveMcpJevP.

%% file: tables/escalation.tex
\begin{tabular}{lrrrrrrrrr}
\toprule
Model & XSS & Trav. & Cmd. & SQLi & L4Sh. & XXE & ShSh. & Benign & Exploits \\
\midrule
Jev & 100 & 99 & 98 & 99 & 100 & 100 & 100 & 5.2 & 98 \\
Clef & 100 & 98 & 98 & 99 & 99 & 100 & 100 & 4.5 & 97 \\
Clef-flash & 70 & 76 & 74 & 68 & 81 & 97 & 91 & 5.1 & 92 \\
\midrule
Qwen2.5-7B & 88 & 82 & 89 & 90 & 85 & 94 & 96 & 4.9 & 82 \\
\midrule
ModernBERT & 98 & 56 & 98 & 94 & 82 & 100 & 100 & 5.1 & 68 \\
Laya (ft) & 94 & 32 & 97 & 93 & 98 & 100 & 100 & 4.8 & -- \\
n-gram model & 88 & 3 & 86 & 68 & 36 & 94 & 91 & 4.6 & 41 \\
\midrule
CRS 4.30.0 & 100 & 86 & 95 & 81 & 100 & 97 & 87 & 5.0 & 66 \\
CRS 3.3.2 & 100 & 85 & 96 & 88 & 100 & 100 & 96 & 5.3 & 64 \\
Anomaly baseline & 63 & 46 & 56 & 55 & 60 & 94 & 70 & 4.5 & 11 \\
\bottomrule
\end{tabular}

%% file: tables/conformal.tex
\begin{tabular}{lrrrrrrrrr}
\toprule
Model & known & XSS & Trav. & Cmd. & SQLi & L4Sh. & XXE & ShSh. & Load \\
\midrule
Jev & 1.3 & 0 & 2 & 2 & 2 & 2 & 0 & 0 & 0.9 \\
Clef & 3.5 & 1 & 9$^{*}$ & 4 & 4 & 3 & 0 & 0 & 1.5 \\
Clef-flash & 5.9 & 6 & 5 & 5 & 11$^{*}$ & 5 & 0 & 4 & 32.6 \\
\midrule
Qwen2.5-7B & 4.8 & 4 & 9$^{*}$ & 4 & 4 & 8 & 0 & 0 & 16.6 \\
\midrule
ModernBERT & 1.2 & 5 & 65$^{*}$ & 4 & 11$^{*}$ & 43$^{*}$ & 0 & 0 & 0.9 \\
Laya (ft) & 2.3 & 14$^{*}$ & 74$^{*}$ & 6 & 18$^{*}$ & 17$^{*}$ & 0 & 0 & 1.0 \\
n-gram model & 4.7 & 10$^{*}$ & 90$^{*}$ & 18$^{*}$ & 38$^{*}$ & 80$^{*}$ & 6 & 4 & 6.7 \\
\midrule
CRS 4.30.0 & 4.8 & 0 & 17$^{*}$ & 4 & 13$^{*}$ & 0 & 0 & 0 & 7.0 \\
CRS 3.3.2 & 4.5 & 0 & 29$^{*}$ & 3 & 11$^{*}$ & 0 & 0 & 4 & 22.7 \\
\bottomrule
\end{tabular}

%% file: tables/controls.tex
\begin{tabular}{lrrrr}
\toprule
Model & Exploits & Twins & Neutral & After \\
\midrule
Jev & 98 & 100/37 & $-$1.4 & 9 \\
Clef & 97 & 98/27 & $-$1.9 & 8 \\
Clef-flash & 92 & 90/13 & $-$4.7 & 9 \\
Qwen2.5-7B & 82 & 84/2 & $-$1.4 & 7 \\
ModernBERT & 68 & 67/1 & $-$0.5 & 6 \\
n-gram model & 41 & 39/3 & $-$0.4 & 3 \\
CRS 4.30.0 & 66 & 74/2 & $-$0.2 & 6 \\
\midrule
n-gram, temporal & 94 & 87/52 & $-$3.1 & -- \\
ModernBERT, temporal & 99 & 100/89 & 0.0 & -- \\
No values, temporal & 93 & 100/100 & -- & -- \\
\bottomrule
\end{tabular}

%% file: sections/results_adversary.tex
\section{Evasion and Flooding (RQ3)}
\label{sec:adversary}

\begin{table}[t]
\caption{Held-out attacks / benign requests (\%) escalated at each model's clean 5\% thresholds,
unedited (clean) and with a parameter asserting the request is benign (note) or instructing the
classifier (instr.), the assertion appended to the attacked value (value), the payload
percent-encoded (enc.), or about 3,000 characters of padding before it (pad). ModernBERT: its kept
run (Section~\ref{sec:models}).}
\label{tab:battery}
\footnotesize
\setlength{\tabcolsep}{2pt}
\input{tables/battery}
\end{table}

Against the second and third adversaries of Section~\ref{sec:threat}, fixed edits lower Jev's and
Clef's escalation of attacks by at most a few points, while an injected assertion cuts that of
Qwen2.5-7B, and padding that of the supervised detectors, by half or more. A search guided by the
model's probability gets past Jev mostly with requests at its threshold, which Jev often escalates
when asked again; against Clef, which answers repeated calls identically, its successes hold. Padded
benign requests, which Clef nearly always escalates, can fill the slow tier.

\subsection{Fixed edits}
\label{sec:battery}
Jev escalates at least \NumBatJevAttackMin{} of the attacks under every fixed edit, against
\NumBatJevCleanOne{} unedited (Table~\ref{tab:battery}). The assertions, the instruction and padding
raise Clef's escalation from \NumBatClefClean{} to as much as \NumBatClefInstr, but encoding every
character of the payload lowers it to \NumBatClefEnc. Qwen2.5-7B follows the injected text: the
benign assertion lowers its escalation of attacks from \NumBatQwenClean{} to \NumBatQwenNote{} as a
parameter and to \NumBatQwenValue{} appended to the attacked value. The rule set escalates
\NumBatCrsoneClean{} of the attacks clean and under each text edit, and \NumBatCrsoneEnc{} encoded.
Padding lowers the supervised detectors' escalation from \NumBatModernbertClean{} to
\NumBatModernbertPad{} (ModernBERT) and from \NumBatTfidfClean{} to \NumBatTfidfPad{} (n-gram model).
On benign traffic the same edits can raise escalation: Jev and Clef escalate \NumBatJevInstrBenign{}
and \NumBatClefInstrBenign{} of benign requests that carry the instruction (Clef-flash
\NumBatClefflashInstrBenign), and ModernBERT at least \NumBatModernbertBenignMin{} of the benign
requests under every edit.

\subsection{Score-guided search}
\label{sec:adaptive}
Of the \NumAdvJevN{} held-out attacks Jev escalated, the search of Section~\ref{sec:robustness} got
\NumAdvJevSuccess{} past it within \NumAdvJevBudget{} queries (\NumAdvJevAtHundred{} within 100).
Success depends on the initial score, as for prompt injection against Jev's action
choices~\cite{wu2026hijacking}: the search got past Jev for \NumAdvJevLowStart{} of the attacks Jev
first scored below 0.93, \NumAdvJevMidStart{} of those from 0.93 to 0.96 and \NumAdvJevHighStart{} of
those from 0.96. The requests it found score \NumAdvJevFinalPMin{} to \NumAdvJevFinalPMax, just below
the threshold, and Jev's answers vary: asked \NumReqJevRepeats{} more times, Jev escalated them in
\NumReqJevEsc{} of its answers. Of the \NumReqJevN{} requests found, \NumReqJevMajority{} stayed past
Jev in most answers and \NumReqJevAll{} in all; the others passed on a favorable answer.

Clef answered identically when asked again: it lost \NumAdvClefSuccess{} of \NumAdvClefN{} attacks
within \NumAdvClefBudget{} queries, Jev \NumAdvJevSameClef{} of the same attacks, and every request
that got past Clef still did on the repeated calls. Clef-flash and Qwen2.5-7B lost every attack they
escalated, after a median of \NumAdvClefflashMedianQ{} queries each. The successful requests transfer
poorly: none that evaded Qwen2.5-7B evades Jev or Clef, and \NumAdvTransferJevClef{} of those that
evaded Jev evade Clef. The search reconciles our battery with JevAdvBench, in which one appended
opinion flips 12.1\% of Jev's general decisions~\cite{hu2026jevadvbench}: no fixed edit moves Jev's
mean recalibrated score of an attack by more than \NumBatJevShiftAbsMax, while a search accumulates
small moves. Smaller budgets help the search, because their thresholds lie higher: its best request
falls below Jev's 1\% threshold (Figure~\ref{fig:budgets}) for \NumAdvJevPastOne{} of the
\NumAdvJevStartOne{} attacks Jev escalates at that budget, and below its 0.1\% threshold for all
\NumAdvJevStartTenth{} it escalates there. Against Clef's thresholds it succeeds for
\NumAdvClefPastOne{} of \NumAdvClefStartOne{} attacks at 1\% and \NumAdvClefPastTenth{} of
\NumAdvClefStartTenth{} at 0.1\%.

\subsection{Flooding the slow tier}
\label{sec:flood}
The attack-like benign requests of Section~\ref{sec:data} are escalated more often than ordinary
traffic by the hosted models, Qwen2.5-7B and the supervised detectors, from \NumClefHardEsc{} of them
by Clef to \NumTfidfHardEsc{} by the n-gram model, against the budget's 5\%. This is the
false-escalation load such traffic imposes, and it sets the cost of flooding the slow tier with
requests the edge cannot easily refuse (Section~\ref{sec:threat}). A slow tier that handles ten
requests per second fills with about 200 ordinary requests per second, but with \NumClefFlood{}
attack-like ones against Clef, \NumJevFlood{} against Jev and \NumTfidfFlood{} against the n-gram
model. An adversary who can query a public model can keep only the requests it escalates: of the
\NumFloodSelectedN{} that Jev escalates, Clef escalates \NumFloodClefSelMin, against
\NumFloodClefAllMin{} of attack-like requests overall. ModernBERT's seed-17 main run, which the
adversary cannot query, escalates \NumFloodModernbertSelMin{} to \NumFloodModernbertSelMax{} of them
depending on the fold, against \NumFloodModernbertAllMin{} to \NumFloodModernbertAllMax{} overall.

Padding is cheaper still. With about 3,000 characters of opaque parameters added, Clef escalates
\NumBatClefPadBenign{} of benign requests and ModernBERT \NumBatModernbertPadBenign{}
(Table~\ref{tab:battery}). The same slow tier then fills at about \NumFloodPadClef{} padded requests
per second against Clef and \NumFloodPadJev{} against Jev, while the n-gram model and both releases of
the rule set escalate at most \NumBatCrsoldonePadBenign{} of them. An edge rule that refuses requests
of that length would also refuse the \NumPadLongerShare{} of the distinct recorded requests
(Section~\ref{sec:data}) that are longer than the shortest padded one (\NumPadMinChars{} characters).

%% file: tables/battery.tex
\begin{tabular}{lrrrrrr}
\toprule
Model & clean & note & instr. & value & enc. & pad \\
\midrule
Jev & 99/5 & 99/5 & 100/100 & 99/18 & 99/10 & 99/43 \\
Clef & 98/4 & 99/25 & 100/99 & 100/91 & 94/7 & 100/97 \\
Clef-flash & 72/5 & 72/7 & 79/24 & 76/9 & 55/10 & 77/38 \\
\midrule
Qwen2.5-7B & 88/4 & 23/0 & 61/0 & 24/0 & 76/23 & 84/29 \\
\midrule
ModernBERT & 93/5 & 98/56 & 99/77 & 98/54 & 96/80 & 43/50 \\
n-gram model & 58/5 & 91/59 & 98/90 & 88/58 & 57/8 & 15/0 \\
\midrule
CRS 4.30.0 & 95/4 & 95/3 & 95/3 & 95/3 & 91/4 & 95/3 \\
CRS 3.3.2 & 96/5 & 96/4 & 96/6 & 96/5 & 93/5 & 95/5 \\
Anomaly baseline & 53/3 & 53/3 & 53/3 & 75/75 & 53/3 & 100/100 \\
\bottomrule
\end{tabular}

%% file: sections/results_operations.tex
\section{The Cost of Running a Fast Tier (RQ4)}
\label{sec:operations}

A hosted decision model adds a few hundred milliseconds to each request, so it suits triage of a
copy of the traffic better than inline use (Section~\ref{sec:background}). A cascade with the rule set
as cheap first stage calls Jev on a fifth of traffic but escalates fewer of the recent exploits than
Jev alone.

\subsection{Latency and throughput}
\label{sec:latency}
Over the \NumJevCalls{} live calls of our leave-one-family-out experiments, made one at a time, Jev
answered in \NumJevLatMedian{}~ms at the median and \NumJevLatNinetyNine{}~ms at the 99th percentile,
most of it the network round trip of \NumApiRoundTrip{}~ms from our client, which a client closer to
the service would largely avoid. Clef and Clef-flash answered \NumClefCalls{} calls each in
\NumClefLatMedian{} and \NumClefflashLatMedian{}~ms at the median and \NumClefLatNinetyNine{} and
\NumClefflashLatNinetyNine{}~ms at the 99th percentile, over a round trip of \NumClefRoundTrip{}~ms,
almost all of it the service's own (Figure~\ref{fig:tradeoff} in Appendix~\ref{sec:additional}).
Clef-flash, about five times as fast as Clef at the median in Cloudflare's
benchmark~\cite{chen2026clef}, matched it. Locally, on one Apple M5 laptop, ModernBERT answers in
\NumLatModernbertMedian{}~ms at the median and the n-gram model in under a millisecond. With 1 to 16
concurrent clients Jev's throughput rose from \NumApiRpsOne{} to \NumApiRpsSixteen{} requests per
second without a throttled request; a service with ten million requests a day averages about 116 per
second, more at peaks.

\subsection{Cost per request}
\label{sec:cost}
With the five questions of our set, a request of the folds costs Jev \NumJevTokensMean{} input tokens
on average, or \$\NumJevCostPerMillion{} per million requests at the published price, and Clef and
Clef-flash \NumClefTokensMean{} tokens, or \$\NumClefCostPerMillion{} and \$\NumClefflashCostPerMillion{}
per million. A service with ten million requests a day would pay about \$\NumJevCostPerYear{} a year
to put Jev in front of all of its traffic and \$\NumClefCostPerYear{} for Clef. On \NumSingleN{}
requests asked only the primary question, which is all the budget and conformal rules use, Jev reads
\NumSingleTokensOne{} tokens instead of \NumSingleTokensFive, about \$\NumSingleCostPerYear{} a year,
and its answer moves by at most \NumSingleMaxDiff. Our benign requests have at most 1,000
characters; longer ones cost more in proportion to their tokens. Under a budget rule the slow tier
examines a share $b$ of benign traffic, so it costs more than the fast tier once one examination
costs more than $1/b$ fast-tier calls: \NumCostBreakFive{} Jev calls (\NumCostBreakFiveUsd) at a 5\%
budget, \NumCostBreakOne{} (\NumCostBreakOneUsd) at 1\% and \NumCostBreakTenth{}
(\NumCostBreakTenthUsd) at 0.1\%.

\subsection{Cascades}
\label{sec:cascade}
A cascade (Section~\ref{sec:routing}) cuts the fast tier's cost: a cheap first stage screens all
traffic and forwards only the requests it ranks highest to Jev, whose threshold is set so that the
cascade as a whole escalates 5\% of benign calibration traffic (\NumCascadeBenignMin{} to
\NumCascadeBenignMax{} of benign test traffic). Forwarding the top 20\% of traffic, the cascade
escalates \NumCascadeTfidfFwdTwentyTraversal{} of held-out traversal with the n-gram model as first
stage and \NumCascadeModernbertFwdTwentyTraversal{} with ModernBERT (main run, seed 17), against
\NumJevEscTraversal{} for Jev alone. The rule set, trained on no attack data, makes a first stage that
forwards the top 20\% by its anomaly score; the cascade then escalates at least
\NumCascadeCrsoneFwdTwentyMin{} of every held-out family (traversal:
\NumCascadeCrsoneFwdTwentyTraversal) while calling Jev on \NumCascadeCrsoneFwdTwentyCalls{} of benign
traffic. On the recent exploits the same cascade escalates \NumCascadeCveEsc, against
\NumJevCveEsc{} for Jev alone, and \NumCascadeCveAuthEsc{} of the authentication bypasses,
\NumCrsCveAuthZero{} of which the rule set scores zero on every request, like most benign traffic.

\subsection{Slow-tier load at a realistic base rate}
\label{sec:baserate}
Rates measured on balanced test sets translate into very different daily volumes once attacks are
rare~\cite{axelsson2000baserate}. For a service with ten million requests a day, one in a thousand of
them an attack from a held-out family, the traversal fold's rates under the conformal rule send
\NumJevDeployEscalations{} requests a day to the slow tier with Jev as the fast tier and
\NumModernbertDeployEscalations{} with ModernBERT. Jev closes \NumJevDeployMissed{} of the
\NumJevDeployAttacks{} attacks and ModernBERT \NumModernbertDeployMissed. Alerts do not reach the slow
tier, but each benign one is a false alarm: Jev raises \NumJevDeployFalseAlerts{} a day and ModernBERT
\NumModernbertDeployFalseAlerts.

%% file: sections/discussion.tex
\section{Discussion}
\label{sec:discussion}

\subsection{Lessons for operators}
\emph{Evaluate a fast tier on families held out from everything it fits, over several training
runs.} Every detector we trained on the known families escalated far fewer of the same traversal
requests once traversal was held out, and for ModernBERT how many fewer changed from one training
run to the next.

\emph{Treat a supervised detector's confident benign verdicts on new attacks with caution.} Its
recalibration map is learned from the families it already knows, and on a new family it can turn
moderate scores into confident benign verdicts, for a fine-tuned decision model as well as for
ModernBERT.

\emph{Pair exploit tests with benign requests of the same shape.} Detectors trained on older
exploits escalated many benign twins of recent exploits.

\emph{Route by a benign budget.} A conformal rule's guarantee covers only attacks exchangeable with
those it was fitted on. Where the known classes separate, the rule closes what a 1\% budget closes;
where they overlap, its benign load ranged from \NumConfLoadClef{} (Clef) to most of the traffic (the
small open models).

\emph{Test a cascade's first stage on held-out families and on recent exploits}
(Section~\ref{sec:cascade}).

\emph{Assume an adversary can query a public fast tier, and judge evasions on repeated calls.} A
search guided by the model's own probability got held-out attacks past Jev and Clef that fixed edits
did not, but most of its successes against Jev did not hold up when the requests were asked again,
and the requests it found rarely evaded another model.

\emph{Budget per service.} Benign escalation varies by service: of the \NumHostsN{} hosts with at
least \NumHostsMinRequests{} benign test requests, Jev escalated no request of \NumJevHostZeroN{} but
\NumJevHostEscMax{} of the requests of one, whose client polls an endpoint with a stray
\texttt{\{\}} in its query string.

\emph{Expect content-judging fast tiers to miss more of the attacks that look like ordinary use.}
For most fast tiers authentication bypasses were the recent exploits they escalated least: such
requests often carry nothing suspicious, and whether one is legitimate depends on the application's
state.

\emph{Decide what a hosted fast tier may see and what happens when it fails.} A hosted model receives
every request body, sign-ups and shopping carts among them, which raises data-protection questions a
self-hosted model avoids; we saw no throttling up to 16 clients but did not test outages.

\subsection{Limitations}
\label{sec:limitations}

\paragraph{Exposure during pretraining.}
A held-out family is unseen by everything a deployment fits, but perhaps not by a pretrained model:
the payload lists, Log4Shell and Shellshock are public, and none of the providers we test documents
its training data in full. Our results for zero-shot models therefore measure behavior on attack
families absent from the deployment's own data, the position of an operator who fits a fast tier on
local traffic, and may overstate how they would meet an attack class absent from every pretraining
corpus. So far only \NumProN{} exploits postdate the models' release, and they belong to known attack
classes. A replay with a model trained before Log4Shell's disclosure failed the screen we set before
running it, an AUROC of at least \NumTzeroGo{} for cross-site scripting and SQL injection on the
development set: T0~\cite{sanh2022t0} ranks both below benign requests (AUROC \NumTzeroAurocXss{} and
\NumTzeroAurocSqli), which recalibration turns into a constant (Section~\ref{sec:routing}). Jev and
Clef escalate at most \NumHostedDropMax{} points less of encoded payloads than of unedited ones, and
of the exploit classes no payload family covers than of all recent exploits, though only
\NumJevCveEscAuth{} of authentication bypasses.

\paragraph{Families as units.}
Only three families leave enough held-out attacks for main folds, and they differ: ModernBERT gives
\NumModernbertConfMissTraversal{} of held-out traversal a probability below 0.05 but
\NumModernbertConfMissXss{} of cross-site scripting. Our intervals describe these families only, and
the rate on a new family may differ.

\paragraph{Training runs.}
ModernBERT's held-out results vary between training runs with identical data
(Section~\ref{sec:trust}), so the three main runs understate their spread; Laya (ft) was trained
once, and its \NumLayaftConfMissTraversal{} confident misses on traversal may vary as much.

\paragraph{Splits.}
Known-family payloads are split by their text, so encoding variants of one payload can sit in both
training and test and flatter known-family results, and with them the gap to held-out families; a
held-out family is excluded whole and cannot leak this way. Benign requests to the same endpoint with
different values also land in different splits: \NumEndpointInTraining{} of benign test requests
share host, method and path with a training request; splits by payload skeleton and by host would
test how much either matters.

\paragraph{Spliced attacks and injection points.}
Splicing gives every attack a real host, path and parameter names, and the format control shows that
these cues no longer separate the classes. A spliced request still carries one payload in one
parameter of a request recorded for other purposes, while real attacks often arrive in sequences
and with their own headers. The header test covers two families, the exploit test leaves out the
\NumCveExclHeader{} exploits that carry part of the attack in a header the compact form drops, and our
filters, at most 1,000 characters and at least one parameter, leave out long bodies and requests
without parameters. Labels come from construction, so some payloads may be inert in their context.

\paragraph{Benign traffic.}
The main benign traffic comes from one browsing campaign recorded through one browser for the 2026
edition of the open-appsec comparison, a browser's view of many sites, which may differ from one
application's inbound traffic, and our second source, CSIC 2010, is synthetic. The results depend on
the benign traffic: on ECML's synthetic benign requests, which the hosted models find suspicious, Jev
escalates only \NumJevEcmlEscLdap{} of held-out LDAP injection (Appendix~\ref{sec:ecml}).

\paragraph{Small supervised detectors.}
Our supervised detectors are small (the n-gram model, ModernBERT with 149M parameters, Laya (ft) with
421M); we did not test whether a large model fine-tuned on the known families, such as Qwen2.5-7B or
Clef-flash, would also close held-out attacks with confidence.

\paragraph{The adaptive search.}
The search uses one set of edits, at most 300 queries and the model's probability, and judges an
evasion by the fast tier's decision, not by an application; an adversary with more queries or other
edits may do better, and a defender that rate-limits or hides scores may make it harder.

\paragraph{Hosted models change.}
Asked \NumJevRepeatN{} of our requests again, Jev changed at most \NumJevRepeatFlipMax{} of the
routing decisions. Our \NumCanaryN{} fixed canary requests, sent weekly from \NumCanaryFirst{} to
\NumCanaryLast, moved by at most \NumCanaryMeanDiff{} on average. We pinned \texttt{jev-1.13.0}; by
\NumCanaryAliasSwitch{} the unpinned alias returned its successor
\texttt{jev-1.14.0}~\cite{typesafe2026docs}. It escalates \NumDriftEscNew{} of held-out traversal
against \NumDriftEscOld, both at a benign rate of \NumDriftBenignOld, but moved the canary
probabilities by \NumDriftCanaryMeanDiff{} on average and pushed \NumDriftCanaryFlips{} canary
decisions across 0.5, so an operator who follows the alias must recalibrate after each change.
Cloudflare gives no version, so Clef may change unannounced.

%% file: sections/conclusion.tex
\section{Conclusion}
\label{sec:conclusion}

In a two-tier pipeline the fast tier decides alone on everything it does not escalate, so how it
treats attack families it was never fitted on bounds how many new attacks the pipeline can catch.
Holding families out of everything a deployment fits, we found that detectors trained on the known
families gave large parts of some held-out families a confident benign verdict, by amounts that
varied between training runs. Jev and Clef, whose budget decisions the hold-out cannot change,
escalated at least \NumClefEscFloor{} of every held-out family within a 5\% benign budget. Benign
twins showed that detectors trained on older exploits had learned what scanner requests look like,
and a search guided by Jev's probability got \NumAdvJevSuccess{} of held-out attacks past it within
\NumAdvJevBudget{} queries but only \NumReqJevRobustShare{} past it in most of \NumReqJevRepeats{}
repeated calls.

Our protocol can be reused on new fast-tier models, with the data checks of
Appendix~\ref{sec:hygiene}: hold each family out of training, recalibration and thresholds over
several training runs, spend a fixed benign budget, pair exploit tests with benign twins, attack
with fixed and adaptive edits, judge evasions on repeated calls, and count the held-out attacks the
fast tier lets through.

%% file: sections/ethics.tex
\section{Ethical Considerations}
\label{sec:ethics}

This study involved no live systems. Every attack in our experiments comes from public sources: the
payloads of the mgm WAF Payload Collection (a GPL-3.0 repository whose README notes that most
payloads carry their sources' MIT license), as packaged by the open-appsec WAF comparison project, and the exploit requests of the
nuclei templates (MIT). Payloads were inserted as text into recorded
requests, and no request reached a live application: the models scored text, and the rule set ran in
a local container with no application behind it. The out-of-band callback domain of the templates
was replaced by a reserved \texttt{.example} name, and in one control by a subdomain of the request's
own host, as text only. The
benign traffic is a public recording (Apache-2.0), used without cookies or other session headers
(Section~\ref{sec:data}), and our artifact contains no text of the recorded requests. SR-BH 2020 (CC0) serves only
for the label audit, which reports counts and row indices.

Requests, attack payloads included, were sent as text to two hosted services under their
providers' terms, among them the edited attacks of the adaptive search (about \NumAdvJevQueries{}
queries to Jev and \NumAdvClefQueries{} to Clef). TypeSafe's terms assign outputs to the customer and do not restrict benchmarking;
Cloudflare's terms permit benchmark tests whose disclosure includes the information needed to
replicate them, which our artifact provides. We report model versions and dates so that the
providers can reproduce our observations.

Three results could inform an attacker: the held-out families that a fast tier closes with
confidence, the rate of legitimate requests that fills a slow tier, and the share of attacks a
score-guided search gets past a hosted model. They rest on public payloads and on edits of known
kinds; we publish the search code but not the requests it found. The study involved no human
participants.

%% file: sections/openscience.tex
\section{Open Science}
\label{sec:openscience}

The artifact, available at \href{https://anonymous.4open.science/r/escalate-or-close-artifact-3794/}{\nolinkurl{https://anonymous.4open.science/r/}\allowbreak\mbox{\nolinkurl{escalate-or-close-artifact-3794/}}}, contains the
code for every experiment; the analysis plan written before the first leave-one-family-out run and
the log of every later deviation; fold manifests as hashes of each item's text; per-item
predictions of every model on every fold; the cached answers of the zero-shot models keyed by text
hash, without request text; and the scripts that rebuild every table, figure and number in this
paper from these files. Datasets are not redistributed: download scripts fetch the original
releases and verify them against SHA-256 checksums, so the artifact ships no payload data; the
nuclei templates are fetched at two pinned commits, \texttt{c8664a1} of 9 October and \texttt{1f4c3f6}
of 23 October 2026~\cite{projectdiscovery2026nuclei,projectdiscovery2026nucleisecond}, and verified by
a digest over every template. CVE
publication dates come from the NVD API~\cite{nist2026nvdapi}; this product uses data from the NVD
API but is not endorsed or certified by the NVD. Open
models are pinned to Hugging Face revisions with file hashes and Jev to \texttt{jev-1.13.0} (the drift
check to \texttt{jev-1.14.0}); Workers AI serves Clef and Clef-flash by name without a version, and we
called them on 19, 22, 23 and 28 October 2026, Clef again on 30 October to repeat the adaptive
search's requests, and both on 31 October to score every benign calibration and test request.
The rule set runs from pinned container images, with release 3.3.2 verified by its SHA-256.
Re-running the hosted models requires API keys, and Jev's answers to identical calls vary by up to
\NumReqJevMaxDiff{} in probability (Section~\ref{sec:adaptive}); every reported number can be regenerated without a key
from the cached answers and the stored latency measurements.

%% file: sections/appendix_ecml.tex
\section{Results on ECML/PKDD 2007}
\label{sec:ecml}

\begin{table}[h]
\caption{ECML/PKDD 2007: held-out AUROC / share of held-out attacks escalated at a 5\% benign
budget (\%), per held-out family.}
\label{tab:ecml}
\footnotesize
\setlength{\tabcolsep}{2pt}
\input{tables/ecml}
\end{table}

We drop the \NumEcmlMultiType{} attacks labeled with two types, keep server-side includes
(\NumEcmlSsiKept{} requests, too few to hold out) as a known family, and hold out the other
\NumEcmlHeldFamilies{} families (\NumEcmlFamilyMin{} to \NumEcmlFamilyMax{} requests) in turn; each
leaves \NumEcmlHeldOutMin{} to \NumEcmlHeldOutMax{} test requests. On ECML, ModernBERT and the n-gram
model lead (Table~\ref{tab:ecml}). ModernBERT escalates at least
\NumModernbertEcmlEscMin{} and the n-gram model at least \NumTfidfEcmlEscMin{} of every held-out
family, a result that likely reflects the corpus more than the models: most of its attack families share a cue
that its benign requests lack (Appendix~\ref{sec:hygiene}). The hosted decision models read the same benign requests as
suspicious. Zero-shot, Jev scores \NumJevEcmlBenignAboveHalf{} of the benign ECML test requests above
0.5 (median \NumJevEcmlBenignMedian), against \NumJevOaBenignAboveHalf{} of the real benign traffic
of open-appsec (median \NumJevOaBenignMedian), and Clef \NumClefEcmlBenignAboveHalf. A monotone
recalibration does not reorder requests, so it cannot separate attacks from benign requests that a
model scores as high, and Jev, Clef and
Clef-flash escalate only \NumJevEcmlEscLdap, \NumClefEcmlEscLdap{} and \NumClefflashEcmlEscLdap{} of
held-out LDAP injection. ECML thus favors models that learn its benign vocabulary over models that
judge each request by its content, which limits its value as a benchmark for held-out attack
families. The local language model behaves in the same way: it scores \NumQwenEcmlBenignAboveHalf{} of the
benign ECML requests above 0.5 and escalates \NumQwenEcmlEscMin{} to \NumQwenEcmlEscMax{} of each
held-out family, and the small open models escalate \NumOpenEcmlEscMin{} to
\NumOpenEcmlEscMax.

%% file: tables/ecml.tex
\begin{tabular}{lrrrrrr}
\toprule
Model & LDAP & OS cmd. & Trav. & SQLi & XPath & XSS \\
\midrule
Jev & 0.78/10 & 0.97/84 & 0.88/53 & 0.96/80 & 0.91/48 & 0.96/81 \\
Clef & 0.78/15 & 0.92/62 & 0.85/44 & 0.95/78 & 0.95/74 & 0.94/69 \\
Clef-flash & 0.68/14 & 0.82/41 & 0.70/19 & 0.84/44 & 0.74/19 & 0.91/60 \\
\midrule
Qwen2.5-7B & 0.76/7 & 0.87/7 & 0.76/6 & 0.86/7 & 0.81/7 & 0.91/7 \\
Laya & 0.53/6 & 0.48/5 & 0.50/5 & 0.53/5 & 0.53/6 & 0.53/5 \\
von & 0.58/7 & 0.48/10 & 0.51/7 & 0.63/10 & 0.64/9 & 0.53/5 \\
Verdict & 0.39/6 & 0.61/5 & 0.44/5 & 0.34/5 & 0.25/5 & 0.33/5 \\
\midrule
ModernBERT & 1.00/100 & 1.00/99 & 0.95/83 & 0.99/98 & 1.00/100 & 1.00/100 \\
n-gram model & 1.00/99 & 0.97/90 & 0.92/79 & 0.99/96 & 1.00/100 & 0.97/95 \\
\bottomrule
\end{tabular}

%% file: sections/appendix_pitfalls.tex
\section{Five Pitfalls in Web Attack Data}
\label{sec:hygiene}

Each of these pitfalls could have distorted the measurement.

\paragraph{The benchmark's attacks share one template.}
Because the harness sends every payload to \texttt{/?p=} or in a body \texttt{p=}, a classifier given
only the method, Content-Type, Host, path and parameter names separates the
harness's attacks from benign traffic with an AUROC of \NumFormatVendorAuroc. This concerns learned
detectors evaluated on these requests; it need not affect the product comparison the benchmark was
built for. On spliced attacks, which share host, path and parameter names with benign
requests, the same classifier reaches an AUROC of \NumFormatSplicedAuroc.

\paragraph{The harness's encoding is shared across families.}
The payloads keep the harness's percent-encoding, which escapes characters that browsing traffic
rarely escapes: an escaped parenthesis (\texttt{\%28}) appears in \NumEscParenBenign{} of benign test
requests but in \NumEscParenMin{} to \NumEscParenMax{} of the cross-site scripting, command-execution,
SQL-injection and Shellshock payloads, and in \NumEscParenTraversal{} of the traversal payloads, whose
escapes include overlong byte sequences such as \texttt{\%c0\%af}. No spliced value contains a
character a browser always escapes, and no benign training value contains \texttt{../}, even
decoded. A character n-gram model reads the shared escapes as attack evidence: refitted on requests
whose query and form values are percent-decoded, it escalates \NumEncNgramXssDec{} of held-out
cross-site scripting instead of \NumEncNgramXssOrig{} and \NumEncNgramCmdexeDec{} of command execution
instead of \NumEncNgramCmdexeOrig, while ModernBERT's results barely change
(Section~\ref{sec:trust}).

\paragraph{SR-BH 2020 labels a vulnerability scan as normal.}
Of the \NumSrbhDistinct{} distinct requests of the SR-BH 2020 honeypot corpus~\cite{riera2022srbh},
\NumSrbhOneClient{} come from one client whose traffic includes an OWASP ZAP active scan; of its
requests labeled normal, \NumSrbhOneClientNormalSigShare{} carry an attack payload, against
\NumSrbhOtherNormalSigShare{} of the other clients'. The mislabeling that Bitussi and
Doriguzzi-Corin estimate at about 13.8\%~\cite{bitussi2026xwad} concentrates there, and we do not
evaluate on SR-BH 2020.

\paragraph{ECML/PKDD 2007's attack families share a cue.}
In \NumEcmlSpaceRunFamilies{} of its six held-out families, \NumEcmlSpaceRunMin{} to
\NumEcmlSpaceRunMax{} of the attacks contain a run of three or more encoded spaces (\texttt{+++}),
against \NumEcmlSpaceRunBenign{} of benign requests, so a model trained on the known families can
flag a held-out one by cues they share: the n-gram model detects each held-out family with an AUROC of
\NumEcmlTfidfNovelAurocMin{} to \NumEcmlTfidfNovelAurocMax. Its benign requests in turn carry
attack-like noise, an encoded quote in \NumEcmlBenignQuote{} of them, which a model that judges
content finds suspicious (Appendix~\ref{sec:ecml}). We report ECML for comparison with prior work and
draw our conclusions from open-appsec.

\paragraph{Scanner templates do not map one to one onto exploit requests.}
A nuclei template is often a probe followed by the exploit, builds its payload with helper functions
such as \texttt{base64}, or only fetches a page to read a version string. An earlier conversion of
ours let a visible probe stand in for an exploit step the compact form hides, replaced helper calls by
random text and counted plain page requests as attacks; on it, Jev escalated \NumCveNaiveJevEsc{} of
the exploits within a 5\% benign budget and its conformal rule closed \NumCveNaiveJevClosed, against
\NumJevCveEsc{} and \NumJevCveClosed{} on the faithful conversion.

%% file: sections/appendix_additional.tex
\section{Additional Results}
\label{sec:additional}

\paragraph{Attack families.}
\begin{table}[htbp]
\caption{open-appsec attack families: distinct payloads (cross-site scripting and path traversal
capped by hash), held-out attacks in the test split of the family's own fold, and the fold's role.
Benign requests: \NumOaBenignTrain{} for training, \NumOaBenignCal{} each for calibration and test.}
\label{tab:data}
\footnotesize
\setlength{\tabcolsep}{4pt}
\begin{tabular}{lrrl}
\toprule
Family & Payloads & Held-out attacks & Fold \\
\midrule
Cross-site scripting & \NumOaPayloadCap & \NumOaHeldOutXss & main \\
Path traversal & \NumOaPayloadCap & \NumOaHeldOutTraversal & main \\
Command execution & \NumOaPayloadsCmdexe & \NumOaHeldOutCmdexe & main \\
SQL injection & \NumOaPayloadsSqli & \NumOaHeldOutSqli & case study \\
Log4Shell & \NumOaPayloadsLogshell & \NumOaHeldOutLogshell & case study \\
XML external entities & \NumOaPayloadsXxe & \NumOaHeldOutXxe & case study \\
Shellshock & \NumOaPayloadsShellshock & \NumOaHeldOutShellshock & case study \\
\bottomrule
\end{tabular}
\end{table}
Table~\ref{tab:data} lists the open-appsec families and their sizes. Spliced attacks are not held to
the benign length cap: \NumLongAttackXxe{} of the held-out XML external entities and
\NumLongAttackMin{} to \NumLongAttackMax{} of each other family's held-out attacks exceed 1,000
characters, and leaving them out moves no model's held-out escalation by more than
\NumLengthEscMax{} points.

\paragraph{Traversal by payload form.}
\begin{table}[htbp]
\caption{Held-out path-traversal attacks (traversal fold) by payload form: an example as
transmitted, the number of held-out attacks, the share ModernBERT gives a recalibrated probability
below 0.05 (MB, \%, mean over its main runs), and the shares Jev and Qwen2.5-7B escalate within a 5\% benign
budget (\%). The example is the shortest payload of its form that ModernBERT gives a probability
below 0.05 in every main run.}
\label{tab:forms}
\footnotesize
\setlength{\tabcolsep}{3pt}
\input{tables/forms}
\end{table}
Table~\ref{tab:forms} breaks ModernBERT's confident misses on held-out traversal down by the
form of the payload. Plain dot-dot sequences are missed as well:
ModernBERT gives \NumFormsPlainModernbertMiss{} of them a probability below 0.05, while Jev escalates
at least \NumFormsJevEscMin{} of every form.

\paragraph{Probabilities as they come and placements in headers.}
\begin{table}[htbp]
\caption{Each model's own probability before recalibration: expected calibration error and
confident misses (Section~\ref{sec:metrics}) on known and held-out (held) families, and the share
of benign requests given $p>0.5$. Open-appsec, mean over the main folds and, for ModernBERT, its
main runs.}
\label{tab:raw}
\small
\setlength{\tabcolsep}{3.5pt}
\input{tables/raw}
\end{table}
\begin{table}[htbp]
\caption{Held-out Log4Shell and Shellshock attacks (\%) escalated by placement in requests with
headers: in a parameter (par.), as User-Agent (UA) or in an added X-Api-Version header (XAV), at a
5\% budget set on \NumHeaderBenign{} benign requests in the same form; Benign: realized rate on as
many benign test requests (\%).}
\label{tab:headers}
\footnotesize
\setlength{\tabcolsep}{3pt}
\input{tables/headers}
\end{table}
Tables~\ref{tab:raw} and~\ref{tab:headers} give the results of Sections~\ref{sec:trust} and~\ref{sec:replay} in full.
Within a 5\% budget set on requests with headers, release 4.30.0 of the rule set escalates only
\NumHdrCrsoneLogshellXapi{} to \NumHdrCrsoneLogshellUa{} of the Log4Shell payloads in a header: most
score exactly 5, the threshold, and the tie-breaker splits them. Jev and Clef escalate the header
placements at realized benign rates of \NumHdrJevBenign{} and \NumHdrClefBenign; Qwen2.5-7B escalates
less in a header than in a parameter, and the anomaly baseline, which models parameter values only,
almost none in a header.
For two requests of the battery and the header test, both over 4,000 characters, none of
Qwen2.5-7B's 20 most likely first tokens was yes or no; they are left out of its rates.

\paragraph{ModernBERT across training runs.}
\begin{table}[htbp]
\caption{ModernBERT's held-out traversal and Log4Shell attacks escalated at a 5\% benign budget /
given a probability below 0.05 (\%), per training run: the main runs (seeds 17, 23 and 42), the kept run (seed 17, kept), a further
retraining of seed 17 (repeat), and further seeds. Every run uses the same
data and settings.}
\label{tab:runs}
\footnotesize
\setlength{\tabcolsep}{4pt}
\input{tables/runs}
\end{table}
Table~\ref{tab:runs} lists every run. In the main runs the traversal fold's 5\% budget threshold lies below 0.05
(\NumModernbertTravTauMin{} to \NumModernbertTravTauMax), so \NumModernbertTravBothMin{} to
\NumModernbertTravBothMax{} of held-out traversal attacks are escalated although they are confident
misses. On the traversal fold, known-family escalation stays between
\NumMbRunKnownEscTraversalMin{} and \NumMbRunKnownEscTraversalMax. An ensemble that averages all runs' probabilities
escalates \NumEnsEscTraversal{} of held-out traversal within the budget, \NumEnsUnionEscTraversal{} with
the runs' disagreement as a second signal, and \NumEnsEscLogshell{} of Log4Shell, yet still gives
\NumEnsConfMissTraversal{} of traversal a probability below 0.05.

\paragraph{Small budgets.}
\begin{table}[htbp]
\caption{Share of held-out attacks (\%) escalated at a 0.1\% benign budget whose threshold is set on
all \NumPoolCal{} benign calibration requests; Benign: realized share of the \NumPoolTest{} benign
test requests escalated (\%). ModernBERT: its kept run.}
\label{tab:small}
\footnotesize
\setlength{\tabcolsep}{3pt}
\input{tables/small}
\end{table}
Table~\ref{tab:small} gives the 0.1\% budget of Section~\ref{sec:small} per family.

\paragraph{The temporal test by attack class.}
\begin{table}[htbp]
\caption{Recent exploits escalated (\%) by attack class at the temporal test's 5\% budget, by the
n-gram model and ModernBERT (MB) trained on the seven families alone (7 fam.) or with the 2015--2024
exploits as well (+old).}
\label{tab:temporalclasses}
\footnotesize
\setlength{\tabcolsep}{3pt}
\input{tables/temporal_classes}
\end{table}
With the older exploits in training both detectors escalate nearly every class
(Table~\ref{tab:temporalclasses}), authentication bypasses included for ModernBERT. For server-side
request forgeries the n-gram model learned the generator's callback domain, which the older exploits share: under
the neutral rendering, with the callback under the request's own host, it escalates
\NumTemporalTfidfSsrfNeutral{} of them instead of \NumTemporalTfidfSsrfOrig.

\paragraph{The small open decision models.}
Table~\ref{tab:smallopen} gives the results of Laya, von and Verdict, used zero-shot with our
questions at Hugging Face revisions \texttt{7b928d8}, \texttt{498ceba} and \texttt{8af2496}. For von we run one forward pass per question with its chain-of-options controller
disabled, which on the development set never changed an answer; for Verdict we take
P(true)/(P(true)+P(false)) from the full output distribution, because its conditional field is
empty whenever the model abstains. The payload of \NumTruncVerdictAll{} of held-out attacks ends
beyond Verdict's input window, and Laya and von cut none. They escalate little more of a held-out family than the budget's share of benign traffic,
and where their classes overlap their conformal rules escalate or alert on most benign traffic.

\paragraph{The conformal guarantee on known families.}
Every model but Laya and Clef-flash closes at most \NumKnownClosedMaxRest{} of known-family attacks
on average and in no fold significantly more than 5\% (exact binomial $p\geq\NumKnownClosedPMinRest$).
Laya closes \NumLayaKnownClosed{} (\NumLayaKnownClosedMin{} to \NumLayaKnownClosedMax{} per fold,
$p$ from \NumLayaKnownClosedPMin{} to \NumLayaKnownClosedPMax) and Clef-flash
\NumClefflashKnownClosed{} (\NumClefflashKnownClosedMin{} to \NumClefflashKnownClosedMax, $p$ from
\NumClefflashKnownClosedPMin{} to \NumClefflashKnownClosedPMax). The guarantee holds on average over
draws of the calibration split, and the folds share most of their calibration attacks, so they are
not independent checks of it.

\paragraph{The rule set at every paranoia level.}
\begin{table}[htbp]
\caption{The CRS at paranoia levels 1 to 4: share of held-out attacks (\%) escalated at a 5\% benign
budget, and the realized benign escalation (\%).}
\label{tab:crslevels}
\footnotesize
\setlength{\tabcolsep}{2.5pt}
\input{tables/crs_levels}
\end{table}
Each level adds rules that match benign requests as well as attacks, so within a budget the higher
levels mostly escalate less (Table~\ref{tab:crslevels}). As shipped, blocking at an anomaly score of 5, release
4.30.0 blocks \NumCrsoneDefBenign{} of benign test requests at level 1 and \NumCrstwoDefBenign{} at
level 2, \NumCrsoneDefHeldMin{} to \NumCrsoneDefHeldMax{} of each held-out family at level 1, and
\NumCrsoneDefCve{} of the recent exploits; release 3.3.2 blocks \NumCrsoldoneDefBenign{} of benign
requests and \NumCrsoldoneDefHeldMin{} to \NumCrsoldoneDefHeldMax{} of each family at level 1. In
requests with headers, release 4.30.0 at level 1 blocks \NumCrsoneDefHdrBenign{} of benign requests.

\paragraph{Other questions for the small open models.}
With any of the five questions of our set, Laya, von and Verdict reach held-out AUROC at most
\NumOpenQuestionAurocMax{} on the main folds, and Jev at least \NumJevQuestionAurocMin. Laya's card
presents it as a base to specialize and warns that its yes-or-no answers can follow their option
labels~\cite{laya2026card}; its answer to our choice question ranks held-out attacks only slightly
better than its yes-or-no answer (AUROC \NumLayaChoiceAurocMin{} to \NumLayaChoiceAurocMax, against
\NumLayaYesnoAurocMin{} to \NumLayaYesnoAurocMax, on the main folds). Verdict
says true more readily for benign requests whatever the proposition, so its answer to whether a
request is legitimate points the right way, but weakly (AUROC \NumVerdictLegitAurocMin{} to
\NumVerdictLegitAurocMax).

\paragraph{Recent exploits by attack class.}
\begin{table}[htbp]
\caption{Exploits for vulnerabilities from 2025 and 2026 by attack class: share (\%) escalated within
a 5\% benign budget / share closed by the conformal rule, averaged over the seven folds
(Section~\ref{sec:data}). MB: ModernBERT's kept run; CRS: 4.30.0, level 1.}
\label{tab:cve}
\footnotesize
\setlength{\tabcolsep}{2.5pt}
\input{tables/cve}
\end{table}
Table~\ref{tab:cve} gives the exploit test of Section~\ref{sec:cve} by attack class.

\paragraph{Latency against escalation.}
\begin{figure}[htbp]
\includegraphics[width=\columnwidth]{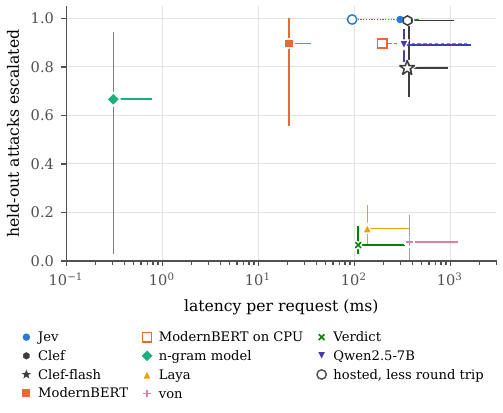}
\Description{Scatter plot with one marker per model: latency of one request on a logarithmic
horizontal axis and the mean share of held-out attacks escalated on the vertical axis, with a
horizontal line to the 99th-percentile latency and a vertical line spanning the seven families. Each
hosted model has a second, hollow marker further left at its latency without the network round
trip.}
\caption{Latency of one request against the share of held-out attacks escalated within a 5\% benign
budget (open-appsec). Markers give the mean over the seven held-out families and vertical lines their
range; horizontal lines run from the median latency to the 99th percentile. Hosted models: one
client, network round trip included (filled) and removed (hollow). Local models:
one request at a time on an Apple M5 laptop (10-core GPU, 24~GB), the n-gram model and the hollow
ModernBERT marker on its CPU; ModernBERT: mean over its main runs. Decision models answer the five
questions of our set, Qwen2.5-7B one.}
\label{fig:tradeoff}
\end{figure}
Figure~\ref{fig:tradeoff} sets each fast tier's latency (Section~\ref{sec:operations}) against its held-out escalation.

\paragraph{Recent exploits.}
Jev escalates \NumJevCveEsc{} of the exploits and its conformal rule closes \NumJevCveClosed{}
(95\% Wilson intervals \NumJevCveEscLo{} to \NumJevCveEscHi{} and \NumJevCveClosedLo{} to
\NumJevCveClosedHi{} over the \NumCveVisible{} exploits, each exploit's outcome averaged over the
seven folds); counting only the requests that carry attack content, which leaves out scanner steps
such as fetching a page, it escalates \NumJevCveContentEsc. For each exploit its conformal rule closes, Jev gives the most suspicious request a
probability between \NumCveJevClosedPMin{} and \NumCveJevClosedPMax. The small open models escalate between
\NumOpenCveEscMin{} and \NumOpenCveEscMax{} of the exploits, and Qwen2.5-7B escalates
\NumQwenCveNovelClassEsc{} of those in classes that no payload family covers. Within an attack
class, Jev's escalated share changes by \NumJevCveTrendEsc{} points per year of publication date
(permutation $p=\NumJevCveTrendEscP$, uncorrected), a test with little power because Jev leaves only
about \NumCveJevNotEscN{} exploits unescalated; it escalated all \NumJevCveAfterN{} exploits
published after its release, all among the \NumProN{} of Section~\ref{sec:robustness}. ModernBERT, released before every dated exploit, changes by
\NumModernbertCveTrendEsc{} points per year ($p=\NumModernbertCveTrendEscP$) and Qwen2.5-7B by
\NumQwenCveTrendEsc{} ($p=\NumQwenCveTrendEscP$), so a date trend reflects how exploits change over
time as well as what a model may have seen.

\paragraph{Disagreement and unions of two models.}
For every zero-shot decision model but Laya, the spread among the four yes-or-no answers separates
held-out attacks from benign requests with an AUROC between \NumSpreadAucBelowMin{} and
\NumSpreadAucBelowMax, below chance; for Laya it points the right way, but weakly (AUROC
\NumLayaSpreadAucMin{} to \NumLayaSpreadAucMax). Across the \NumUnionPairs{} model pairs (Jev with the n-gram model or ModernBERT's seed-17 main run, and
Laya with the n-gram model) and seven held-out families, the
smallest exact McNemar $p$ for a gain of a union over its better model was \NumUnionBestGainP.

\paragraph{Other recalibration methods.}
\begin{table}[htbp]
\caption{Confident misses (\%) on known / held-out families, mean over the main open-appsec
folds, with each model's own probability (none), Platt scaling (used throughout), temperature
scaling and isotonic regression, each fitted on the calibration split at equal class mass.
ModernBERT: its main runs. $^{*}$: the map leaves a calibration error above 0.1 on known
families.}
\label{tab:recal}
\footnotesize
\setlength{\tabcolsep}{3pt}
\input{tables/recal}
\end{table}
Table~\ref{tab:recal} gives the confident misses under each map.

\paragraph{A second benign source.}
\begin{table}[htbp]
\caption{CSIC 2010 normal traffic as the benign source (\NumCsicBenignCal{} calibration and
\NumCsicBenignTest{} test requests): share of held-out attacks (\%) escalated at a 5\% benign budget
per family, and the realized benign rate (\%). ModernBERT: one run (seed 17) trained on the CSIC
folds.}
\label{tab:csic}
\footnotesize
\setlength{\tabcolsep}{3pt}
\input{tables/csic}
\end{table}
Table~\ref{tab:csic} gives the CSIC 2010 test of Section~\ref{sec:budget}. Clef-flash escalates at least \NumCsicClefflashEscFloor{} of every held-out
family. ModernBERT's run, trained on the CSIC folds, gives \NumCsicModernbertConfMissTraversal{} of
held-out traversal a probability below 0.05, and the n-gram model \NumCsicTfidfConfMissTraversal.
Qwen2.5-7B, which scores \NumCsicQwenBenignAboveHalf{} of CSIC's benign requests above 0.5, escalates
only \NumCsicQwenEscMin{} to \NumCsicQwenEscMax{} of each family.

\paragraph{A novelty-detection baseline.}
\begin{table}[htbp]
\caption{ModernBERT's kept run with a Mahalanobis novelty score: held-out attacks (\%) escalated at
a 5\% benign budget by the attack probability alone, the novelty score alone, and either of the two,
their thresholds sharing the budget; Benign: realized rate of the last rule (\%); AUROC: of the
novelty score, held-out attacks against benign test requests.}
\label{tab:novelty}
\footnotesize
\setlength{\tabcolsep}{3pt}
\input{tables/novelty}
\end{table}
The novelty score is the smallest Mahalanobis distance of the input to the run's classification
layer from the means of its training classes, benign traffic and each known family, with one shared
covariance (Table~\ref{tab:novelty}).

\paragraph{The primary comparisons in full.}
\begin{table}[!htbp]
\caption{AUROC, calibration error and confident misses after recalibration
(Section~\ref{sec:metrics}) on known and held-out (held) families, open-appsec, mean over the main
folds and, for ModernBERT, its three main runs. $^{\dagger}$: as in Table~\ref{tab:trust}.}
\label{tab:levels}
\small
\setlength{\tabcolsep}{3.5pt}
\input{tables/levels}
\end{table}

Table~\ref{tab:levels} gives the levels behind Table~\ref{tab:trust}, Table~\ref{tab:primary} the
primary comparisons of every model with P3, and Table~\ref{tab:primaryfamily} P1 and P2 for each main
family alone. Traversal accounts for most of each supervised detector's growth in both, while the
hosted models' confident-miss rates change by at most \NumHostedFamPTwoAbsMax{} points on any main
family. Resampling payload skeletons instead of requests (Section~\ref{sec:metrics}) widens the P2
interval of ModernBERT's seed-17 main run to \NumClusterModernbertPTwoLo{} to
\NumClusterModernbertPTwoHi{} points, still excluding zero for every supervised detector.

\begin{table*}[tp]
\caption{P1 and P2 for each main held-out family alone: held-out minus known calibration error and
confident-miss rate (percentage points) after recalibration, with 95\% intervals from the stratified
Bayesian bootstrap, neither adjusted nor tested. Symbols as in Table~\ref{tab:primary}.}
\label{tab:primaryfamily}
\footnotesize
\setlength{\tabcolsep}{3pt}
\input{tables/primary_family}
\end{table*}

\begin{table*}[tp]
\caption{Primary comparisons on the main open-appsec folds (Section~\ref{sec:metrics}) for every
model and ModernBERT's three main runs: P1 and P2 as in Table~\ref{tab:trust}, with unadjusted 95\%
intervals, and P3, the held-out attacks gained / lost by escalating on confidence or disagreement
against confidence alone at an equal 5\% budget, with the Holm-adjusted $p$ of the exact McNemar
test. $^{\dagger}$, $^{\ddagger}$: as in Table~\ref{tab:trust}; $^{\S}$: outside the analysis plan,
not tested. --: P3 needs
the four yes-or-no answers of a decision model (Qwen2.5-7B answers one question, the supervised
detectors none); Verdict is recalibrated to a constant.}
\label{tab:primary}
\small
\input{tables/primary}
\end{table*}

\begin{table*}[tp]
\caption{The four exploits described in the text, closed by Jev's conformal rule in every fold: the
exploit's request that Jev scores highest (Host header omitted), Jev's probability for it, and the
outcome under the 5\% budget rule, the same in every fold.}
\label{tab:cveexamples}
\footnotesize
\setlength{\tabcolsep}{4pt}
\input{tables/cve_examples}
\end{table*}

\begin{table*}[tp]
\caption{The small open decision models, zero-shot: held-out AUROC over the seven folds (lowest to
highest), held-out attacks escalated at a 5\% benign budget per family (\%), the realized benign
rate (\%), the recent exploits escalated (\%), and the conformal rule's benign load (escalated or
alerted, main folds, \%).}
\label{tab:smallopen}
\footnotesize
\setlength{\tabcolsep}{4pt}
\input{tables/small_open}
\end{table*}

%% file: tables/forms.tex
\begin{tabular}{llrrrr}
\toprule
Form & Example & $n$ & MB & Jev & Qwen \\
\midrule
Plain \texttt{../} & \texttt{../\allowbreak{}/\allowbreak{}/\allowbreak{}boot.ini} & 790 & 76 & 100 & 90 \\
\texttt{\%u} escapes & \texttt{\%3F.\%uF025boot.ini} & 415 & 69 & 99 & 78 \\
Overlong UTF-8 & \texttt{..\%c0\%af..\%c0\%afboot.ini} & 552 & 34 & 99 & 78 \\
URL-encoded & \texttt{0x2e0x2e\%5Cetc\%5Cissue} & 246 & 39 & 100 & 86 \\
Other encodings & \texttt{..\%c0\%qfboot.ini} & 855 & 49 & 99 & 78 \\
\bottomrule
\end{tabular}

%% file: tables/raw.tex
\begin{tabular}{lrrrrr}
\toprule
 & \multicolumn{2}{c}{ECE} & \multicolumn{2}{c}{Conf.\ miss (\%)} & Benign \\
\cmidrule(lr){2-3}\cmidrule(lr){4-5}
Model & known & held & known & held & $p>0.5$ (\%) \\
\midrule
Jev & 0.08 & 0.08 & 0.0 & 0.0 & 0.6 \\
Clef & 0.07 & 0.07 & 0.4 & 0.4 & 1.7 \\
Clef-flash & 0.19 & 0.19 & 4.6 & 4.0 & 2.7 \\
\midrule
Qwen2.5-7B & 0.08 & 0.07 & 4.6 & 4.9 & 11.7 \\
\midrule
ModernBERT & 0.01 & 0.18 & 1.7 & 30.0 & 0.2 \\
Laya (ft) & 0.02 & 0.18 & 3.3 & 32.9 & 0.4 \\
n-gram model & 0.13 & 0.42 & 14.4 & 55.3 & 0.0 \\
\bottomrule
\end{tabular}

%% file: tables/headers.tex
\begin{tabular}{lrrrrrrr}
\toprule
 & & \multicolumn{3}{c}{Log4Shell} & \multicolumn{3}{c}{Shellshock} \\
\cmidrule(lr){3-5}\cmidrule(lr){6-8}
Model & Benign & par. & UA & XAV & par. & UA & XAV \\
\midrule
Jev & 4.4 & 100 & 100 & 100 & 100 & 100 & 100 \\
Clef & 4.3 & 99 & 100 & 100 & 100 & 100 & 100 \\
Clef-flash & 4.4 & 75 & 90 & 86 & 83 & 74 & 61 \\
\midrule
Qwen2.5-7B & 4.5 & 86 & 65 & 73 & 96 & 35 & 52 \\
\midrule
CRS 4.30.0 & 5.1 & 100 & 67 & 66 & 100 & 83 & 65 \\
CRS 3.3.2 & 4.0 & 100 & 3 & 2 & 96 & 96 & 96 \\
Anomaly baseline & 4.6 & 60 & 4 & 4 & 70 & 0 & 0 \\
\bottomrule
\end{tabular}

%% file: tables/runs.tex
\begin{tabular}{lrr}
\toprule
Run & Traversal & Log4Shell \\
\midrule
seed 17 & 53/59 & 74/42 \\
seed 23 & 61/53 & 89/32 \\
seed 42 & 53/55 & 82/39 \\
seed 17, kept & 81/37 & 96/18 \\
seed 17, repeat & 55/52 & -- \\
seed 1 & 79/37 & 96/12 \\
seed 2 & 65/53 & 88/20 \\
seed 3 & 50/60 & 60/51 \\
seed 4 & 67/45 & -- \\
seed 5 & 56/56 & -- \\
\bottomrule
\end{tabular}

%% file: tables/small.tex
\begin{tabular}{lrrrrrrrr}
\toprule
Model & Benign & XSS & Trav. & Cmd. & SQLi & L4Sh. & XXE & ShSh. \\
\midrule
Jev & 0.08 & 86 & 82 & 88 & 92 & 92 & 97 & 91 \\
Clef & 0.10 & 78 & 68 & 84 & 83 & 82 & 89 & 91 \\
Clef-flash & 0.10 & 10 & 10 & 8 & 20 & 11 & 63 & 30 \\
\midrule
ModernBERT & 0.09 & 68 & 6 & 94 & 79 & 31 & 97 & 100 \\
n-gram model & 0.11 & 50 & 0 & 76 & 54 & 4 & 94 & 83 \\
\midrule
CRS 4.30.0 & 0.09 & 96 & 49 & 46 & 47 & 100 & 60 & 65 \\
CRS 3.3.2 & 0.07 & 94 & 37 & 86 & 35 & 11 & 86 & 74 \\
Anomaly baseline & 0.09 & 39 & 32 & 33 & 30 & 30 & 63 & 43 \\
\bottomrule
\end{tabular}

%% file: tables/temporal_classes.tex
\begin{tabular}{lrrrrr}
\toprule
Attack class & $n$ & n-gram, 7 fam. & n-gram, +old & MB, 7 fam. & MB, +old \\
\midrule
File inclusion & 89 & 58 & 97 & 93 & 100 \\
Code execution & 85 & 15 & 95 & 61 & 100 \\
XSS & 67 & 72 & 99 & 97 & 100 \\
SQL injection & 56 & 29 & 93 & 91 & 98 \\
File upload & 44 & 23 & 98 & 48 & 98 \\
Auth.\ bypass & 35 & 6 & 71 & 17 & 94 \\
SSRF & 35 & 3 & 94 & 17 & 100 \\
Other & 12 & 0 & 100 & 58 & 100 \\
\bottomrule
\end{tabular}

%% file: tables/crs_levels.tex
\begin{tabular}{lrrrrrrrr}
\toprule
Release, level & XSS & Trav. & Cmd. & SQLi & L4Sh. & XXE & ShSh. & Benign \\
\midrule
4.30.0, 1 & 100 & 86 & 95 & 81 & 100 & 97 & 87 & 5.0 \\
4.30.0, 2 & 92 & 41 & 77 & 73 & 100 & 100 & 65 & 4.5 \\
4.30.0, 3 & 81 & 27 & 68 & 61 & 89 & 100 & 57 & 4.4 \\
4.30.0, 4 & 32 & 17 & 41 & 37 & 53 & 89 & 35 & 4.8 \\
\midrule
3.3.2, 1 & 100 & 85 & 96 & 88 & 100 & 100 & 96 & 5.3 \\
3.3.2, 2 & 95 & 39 & 85 & 81 & 12 & 91 & 70 & 4.6 \\
3.3.2, 3 & 78 & 29 & 76 & 63 & 9 & 97 & 57 & 4.7 \\
3.3.2, 4 & 29 & 16 & 42 & 37 & 12 & 77 & 26 & 4.8 \\
\bottomrule
\end{tabular}

%% file: tables/cve.tex
\begin{tabular}{lrrrrrrr}
\toprule
Attack class & $n$ & Jev & Clef & Qwen & MB & n-gram & CRS \\
\midrule
File inclusion & 89 & 100/0 & 100/0 & 97/1 & 85/25 & 67/31 & 92/9 \\
Code execution & 85 & 96/5 & 94/7 & 82/6 & 65/43 & 28/72 & 56/40 \\
XSS & 67 & 100/0 & 100/0 & 96/0 & 93/15 & 74/21 & 94/6 \\
SQL injection & 56 & 100/2 & 98/4 & 95/0 & 90/14 & 34/66 & 91/8 \\
File upload & 44 & 100/11 & 100/3 & 80/7 & 53/61 & 27/70 & 48/49 \\
Auth.\ bypass & 35 & 86/23 & 86/47 & 31/33 & 15/91 & 14/81 & 11/84 \\
SSRF & 35 & 97/6 & 97/6 & 54/14 & 22/85 & 12/87 & 20/75 \\
Other & 12 & 100/8 & 100/8 & 67/17 & 60/43 & 5/94 & 25/76 \\
\midrule
All & 423 & 98/5 & 97/7 & 82/7 & 68/40 & 41/57 & 66/32 \\
\bottomrule
\end{tabular}

%% file: tables/recal.tex
\begin{tabular}{lrrrr}
\toprule
Model & None & Platt & Temperature & Isotonic \\
\midrule
Jev & 0.0/0.0 & 0.4/0.8 & 0.6/0.9 & 0.5/1.3 \\
Clef & 0.4/0.4 & 0.8/1.1 & 0.9/1.2 & 0.8/1.2 \\
Clef-flash & 4.6/4.0$^{*}$ & 0.9/0.8 & 0.4/0.4$^{*}$ & 1.1/1.0 \\
\midrule
Qwen2.5-7B & 4.6/4.9$^{*}$ & 1.3/1.6 & 0.9/1.0 & 0.9/1.2 \\
\midrule
ModernBERT, seed 17 & 1.9/33.9 & 0.9/21.6 & 1.5/27.8 & 1.1/26.1 \\
\quad seed 23 & 1.6/27.5 & 1.0/19.8 & 1.3/21.9 & 1.0/19.9 \\
\quad seed 42 & 1.7/28.7 & 1.2/20.9 & 1.3/23.3 & 1.2/25.9 \\
Laya (ft) & 3.3/32.9 & 3.3/32.3 & 3.3/32.9 & 1.7/25.2 \\
n-gram model & 14.4/55.3$^{*}$ & 0.9/21.6 & 0.0/3.6$^{*}$ & 1.3/25.5 \\
\bottomrule
\end{tabular}

%% file: tables/csic.tex
\begin{tabular}{lrrrrrrrr}
\toprule
Model & XSS & Trav. & Cmd. & SQLi & L4Sh. & XXE & ShSh. & Benign \\
\midrule
Jev & 100 & 100 & 100 & 100 & 100 & 100 & 100 & 4.1 \\
Clef & 100 & 100 & 100 & 100 & 100 & 100 & 100 & 5.0 \\
Clef-flash & 96 & 94 & 96 & 93 & 93 & 100 & 100 & 5.3 \\
\midrule
Qwen2.5-7B & 36 & 29 & 48 & 47 & 52 & 54 & 52 & 4.8 \\
\midrule
ModernBERT, seed 17 & 100 & 83 & 99 & 100 & 100 & 100 & 100 & 4.9 \\
n-gram model & 100 & 48 & 97 & 95 & 100 & 100 & 100 & 5.5 \\
\midrule
CRS 4.30.0 & 100 & 78 & 94 & 76 & 100 & 91 & 83 & 5.1 \\
CRS 3.3.2 & 100 & 84 & 97 & 88 & 100 & 100 & 96 & 4.8 \\
Anomaly baseline & 99 & 81 & 91 & 85 & 98 & 100 & 100 & 5.0 \\
\bottomrule
\end{tabular}

%% file: tables/novelty.tex
\begin{tabular}{lrrrrr}
\toprule
Held out & Probability & Novelty & Either & Benign & AUROC \\
\midrule
Cross-site scripting & 95 & 87 & 95 & 4.6 & 0.958 \\
Path traversal & 81 & 83 & 87 & 5.0 & 0.948 \\
Command execution & 98 & 43 & 98 & 5.3 & 0.902 \\
SQL injection & 95 & 93 & 95 & 4.8 & 0.974 \\
Log4Shell & 96 & 85 & 92 & 4.7 & 0.946 \\
XML external entities & 100 & 89 & 100 & 5.1 & 0.977 \\
Shellshock & 100 & 22 & 100 & 4.9 & 0.894 \\
\bottomrule
\end{tabular}

%% file: tables/levels.tex
\begin{tabular}{lrrrrrr}
\toprule
 & \multicolumn{2}{c}{AUROC} & \multicolumn{2}{c}{Calib.\ error} & \multicolumn{2}{c}{Conf.\ miss (\%)} \\
\cmidrule(lr){2-3}\cmidrule(lr){4-5}\cmidrule(lr){6-7}
Model & known & held & known & held & known & held \\
\midrule
Jev & 0.998 & 0.997 & 0.00 & 0.02$^{\dagger}$ & 0 & 1 \\
Clef & 0.996 & 0.995 & 0.01 & 0.04$^{\dagger}$ & 1 & 1 \\
Clef-flash & 0.931 & 0.933 & 0.02 & 0.03 & 1 & 1 \\
\midrule
Qwen2.5-7B & 0.971 & 0.971 & 0.04 & 0.05 & 1 & 2 \\
\midrule
ModernBERT & 0.999 & 0.959 & 0.01 & 0.22$^{\dagger}$ & 1 & 21$^{\dagger}$ \\
Laya (ft) & 0.995 & 0.906 & 0.03 & 0.21$^{\dagger}$ & 3 & 32$^{\dagger}$ \\
n-gram model & 0.987 & 0.829 & 0.02 & 0.21$^{\dagger}$ & 1 & 22$^{\dagger}$ \\
\bottomrule
\end{tabular}

%% file: tables/primary_family.tex
\begin{tabular}{lrrrrrr}
\toprule
 & \multicolumn{2}{c}{Cross-site scripting} & \multicolumn{2}{c}{Path traversal} & \multicolumn{2}{c}{Command execution} \\
\cmidrule(lr){2-3}\cmidrule(lr){4-5}\cmidrule(lr){6-7}
Model & P1 & P2 & P1 & P2 & P1 & P2 \\
\midrule
Jev & +0.022 [0.005, 0.025] & $-$0.5 [$-$1.0, $-$0.1] & +0.028 [0.018, 0.035] & +0.5 [$-$0.1, 0.9] & +0.008 [$-$0.005, 0.018] & +1.2 [0.5, 2.1] \\
Clef$^{\ddagger}$ & +0.024 [0.011, 0.040] & $-$0.9 [$-$1.7, $-$0.4] & +0.046 [0.030, 0.055] & +0.9 [0.1, 1.5] & +0.007 [$-$0.006, 0.025] & +1.0 [0.1, 1.9] \\
Clef-flash$^{\ddagger}$ & +0.027 [0.002, 0.044] & $-$0.4 [$-$1.4, 0.3] & +0.002 [$-$0.021, 0.022] & +0.4 [$-$0.1, 0.9] & +0.007 [$-$0.014, 0.027] & $-$0.2 [$-$1.0, 0.5] \\
\midrule
Qwen2.5-7B & +0.005 [$-$0.011, 0.019] & $-$0.4 [$-$1.4, 0.3] & +0.004 [$-$0.012, 0.018] & +1.5 [0.5, 2.5] & +0.015 [0.000, 0.030] & $-$0.1 [$-$1.0, 0.7] \\
Laya & +0.069 [0.048, 0.094] & 0.0 [0.0, 0.0] & $-$0.019 [$-$0.036, $-$0.001] & 0.0 [0.0, 0.0] & +0.029 [0.008, 0.050] & 0.0 [0.0, 0.0] \\
von & +0.012 [$-$0.008, 0.036] & 0.0 [0.0, 0.0] & +0.036 [0.015, 0.061] & 0.0 [0.0, 0.0] & +0.034 [0.014, 0.057] & 0.0 [0.0, 0.0] \\
Verdict & -- & -- & -- & -- & -- & -- \\
\midrule
ModernBERT, seed 17 & +0.132 [0.115, 0.153] & +1.8 [0.8, 2.7] & +0.488 [0.475, 0.505] & +58.1 [56.2, 60.0] & +0.031 [0.017, 0.048] & +2.4 [1.4, 3.5] \\
\quad seed 23$^{\S}$ & +0.119 [0.103, 0.141] & +1.8 [0.8, 2.7] & +0.452 [0.438, 0.470] & +51.8 [49.8, 53.7] & +0.038 [0.024, 0.053] & +2.9 [1.8, 4.2] \\
\quad seed 42$^{\S}$ & +0.124 [0.107, 0.145] & +2.9 [1.7, 4.0] & +0.439 [0.425, 0.456] & +54.2 [52.2, 56.1] & +0.024 [0.010, 0.042] & +2.0 [1.0, 3.1] \\
Laya (ft)$^{\ddagger}$ & +0.120 [0.100, 0.138] & +10.2 [8.3, 12.0] & +0.378 [0.361, 0.393] & +72.4 [70.4, 74.3] & +0.042 [0.028, 0.058] & +4.4 [2.9, 6.1] \\
n-gram model & +0.067 [0.044, 0.085] & +2.8 [1.8, 3.8] & +0.400 [0.382, 0.416] & +51.5 [49.5, 53.4] & +0.102 [0.081, 0.123] & +7.8 [6.2, 9.5] \\
\bottomrule
\end{tabular}

%% file: tables/primary.tex
\begin{tabular}{lrrrr}
\toprule
Model & P1: calib.\ error & P2: conf.\ miss (points) & P3: gained / lost & P3: Holm $p$ \\
\midrule
Jev & +0.019 [0.011, 0.023]$^{\dagger}$ & +0.4 [0.1, 0.8] & 1 / 12 & 0.027 \\
Clef$^{\ddagger}$ & +0.026 [0.018, 0.033]$^{\dagger}$ & +0.3 [$-$0.1, 0.7] & 8 / 79 & $<$0.001 \\
Clef-flash$^{\ddagger}$ & +0.012 [$-$0.003, 0.020] & $-$0.1 [$-$0.5, 0.3] & 29 / 1{,}018 & $<$0.001 \\
\midrule
Qwen2.5-7B & +0.008 [$-$0.001, 0.016] & +0.3 [$-$0.2, 0.8] & -- & -- \\
Laya & +0.026 [0.015, 0.039]$^{\dagger}$ & 0.0 [0.0, 0.0] & 126 / 193 & 0.002 \\
von & +0.027 [0.016, 0.041]$^{\dagger}$ & 0.0 [0.0, 0.0] & 90 / 390 & $<$0.001 \\
Verdict & -- & -- & 27 / 153 & $<$0.001 \\
\midrule
ModernBERT, seed 17 & +0.217 [0.209, 0.228]$^{\dagger}$ & +20.8 [20.0, 21.6]$^{\dagger}$ & -- & -- \\
\quad seed 23$^{\S}$ & +0.203 [0.195, 0.214] & +18.8 [18.0, 19.6] & -- & -- \\
\quad seed 42$^{\S}$ & +0.196 [0.188, 0.207] & +19.7 [18.8, 20.5] & -- & -- \\
Laya (ft)$^{\ddagger}$ & +0.180 [0.170, 0.190]$^{\dagger}$ & +29.0 [28.0, 30.0]$^{\dagger}$ & -- & -- \\
n-gram model & +0.190 [0.177, 0.200]$^{\dagger}$ & +20.7 [19.8, 21.6]$^{\dagger}$ & -- & -- \\
\bottomrule
\end{tabular}

%% file: tables/cve_examples.tex
\begin{tabular}{l>{\raggedright\arraybackslash}p{0.68\textwidth}rc}
\toprule
CVE & Request Jev scores highest (Host omitted) & Jev $p$ & Budget \\
\midrule
CVE-2026-42018 & \texttt{POST /\allowbreak{}access/\allowbreak{}api/\allowbreak{}v1/\allowbreak{}aws/\allowbreak{}token/\allowbreak{} \{\}} & 0.22 & closed \\
CVE-2025-58434 & \texttt{POST /\allowbreak{}api/\allowbreak{}v1/\allowbreak{}account/\allowbreak{}forgot-password \{"user":\allowbreak{}\{"email":\allowbreak{}"admin"\}\}} & 0.38 & esc. \\
CVE-2026-59726 & \texttt{POST /\allowbreak{}mcp \{"jsonrpc":\allowbreak{}"2.0",\allowbreak{}"method":\allowbreak{}"tools/\allowbreak{}call",\allowbreak{}"id":\allowbreak{}1,\allowbreak{}"params":\allowbreak{}\{"name":\allowbreak{}"ruflo\_\_terminal\_execute",\allowbreak{}"arguments":\allowbreak{}\{"command":\allowbreak{}"id"\}\}\}} & 0.17 & closed \\
CVE-2026-53787 & \texttt{GET /\allowbreak{}media/\allowbreak{}amasty\_checkout/\allowbreak{}k/\allowbreak{}x/\allowbreak{}kxmbikzs.txt} & 0.29 & esc. \\
\bottomrule
\end{tabular}

%% file: tables/small_open.tex
\begin{tabular}{lrrrrrrrrrrr}
\toprule
Model & AUROC & XSS & Trav. & Cmd. & SQLi & L4Sh. & XXE & ShSh. & Benign & Exploits & Load \\
\midrule
Laya & 0.56--0.79 & 10 & 7 & 13 & 11 & 16 & 23 & 13 & 4.7 & 9 & 83.7 \\
von & 0.59--0.71 & 10 & 7 & 19 & 7 & 7 & 0 & 4 & 4.8 & 16 & 83.5 \\
Verdict & 0.21--0.44 & 5 & 4 & 5 & 6 & 3 & 14 & 9 & 4.5 & 7 & 95.5 \\
\bottomrule
\end{tabular}